%% file: main.tex
\documentclass[ALICE,manyauthors]{cernphprep}
\usepackage[comma,square,numbers,sort&compress]{natbib}
\usepackage{hyperref}
\usepackage{lineno}
\usepackage{xspace}
\usepackage{xcolor} 
\usepackage{csquotes}
\usepackage{multirow}
\usepackage{upgreek}
\usepackage{booktabs}
\usepackage[T1]{fontenc}
\usepackage{orcidlink}
\begin{document}
\input{commands.tex}

\begin{titlepage}
\PHyear{2024}       
\PHnumber{055}      
\PHdate{23 February}  

\title{Measurement of beauty-quark production in pp collisions at $\mathbf{\sqrt{\textit{s}}=13}$ TeV via non-prompt D mesons}
\ShortTitle{Non-prompt D-meson production in pp collisions at $\s=13$ TeV}   

\Collaboration{ALICE Collaboration\thanks{See Appendix~\ref{app:collab} for the list of collaboration members}}
\ShortAuthor{ALICE Collaboration} 

\input{abstract.tex}
\end{titlepage}

\setcounter{page}{2} 


\input{introduction.tex}
\input{apparatus.tex}
\input{analysis.tex}
\input{systematics.tex}
\input{results.tex}

\input{summary.tex}


\newenvironment{acknowledgement}{\relax}{\relax}
\begin{acknowledgement}
\section*{Acknowledgements}
\input{fa_2024-02-15_Opt_C.tex}
\end{acknowledgement}

\bibliographystyle{utphys}   
\bibliography{bibliography}

\newpage
\appendix

%
%

\section{The ALICE Collaboration}
\label{app:collab}
\input{2024-02-15-Alice_Authorlist_2024-02-15_Opt_C.tex}
\end{document}

%% file: commands.tex
%

\newcommand{\pp}           {pp\xspace}
\newcommand{\ppbar}        {\mbox{$\mathrm {p\overline{p}}$}\xspace}
\newcommand{\XeXe}         {\mbox{Xe--Xe}\xspace}
\newcommand{\PbPb}         {\mbox{Pb--Pb}\xspace}
\newcommand{\pA}           {\mbox{pA}\xspace}
\newcommand{\pPb}          {\mbox{p--Pb}\xspace}
\newcommand{\AuAu}         {\mbox{Au--Au}\xspace}
\newcommand{\dAu}          {\mbox{d--Au}\xspace}

\newcommand{\s}            {\ensuremath{\sqrt{s}}\xspace}
\newcommand{\snn}          {\ensuremath{\sqrt{s_{\mathrm{NN}}}}\xspace}
\newcommand{\pt}           {\ensuremath{p_{\rm T}}\xspace}
\newcommand{\meanpt}       {$\langle p_{\mathrm{T}}\rangle$\xspace}
\newcommand{\ycms}         {\ensuremath{y_{\rm CMS}}\xspace}
\newcommand{\ylab}         {\ensuremath{y_{\rm lab}}\xspace}
\newcommand{\etarange}[1]  {\mbox{$\left | \eta \right |~<~#1$}}
\newcommand{\yrange}[1]    {\mbox{$\left | y \right |~<~#1$}}
\newcommand{\dndy}         {\ensuremath{\mathrm{d}N_\mathrm{ch}/\mathrm{d}y}\xspace}
\newcommand{\dndeta}       {\ensuremath{\mathrm{d}N_\mathrm{ch}/\mathrm{d}\eta}\xspace}
\newcommand{\avdndeta}     {\ensuremath{\langle\dndeta\rangle}\xspace}
\newcommand{\dNdy}         {\ensuremath{\mathrm{d}N_\mathrm{ch}/\mathrm{d}y}\xspace}
\newcommand{\Npart}        {\ensuremath{N_\mathrm{part}}\xspace}
\newcommand{\Ncoll}        {\ensuremath{N_\mathrm{coll}}\xspace}
\newcommand{\dEdx}         {\ensuremath{\textrm{d}E/\textrm{d}x}\xspace}
\newcommand{\RpPb}         {\ensuremath{R_{\rm pPb}}\xspace}

\newcommand{\nineH}        {$\sqrt{s}~=~0.9$~Te\kern-.1emV\xspace}
\newcommand{\seven}        {$\sqrt{s}~=~7$~Te\kern-.1emV\xspace}
\newcommand{\twoH}         {$\sqrt{s}~=~0.2$~Te\kern-.1emV\xspace}
\newcommand{\twosevensix}  {$\sqrt{s}~=~2.76$~Te\kern-.1emV\xspace}
\newcommand{\five}         {$\sqrt{s}~=~5.02$~Te\kern-.1emV\xspace}
\newcommand{\twosevensixnn}{$\sqrt{s_{\mathrm{NN}}}~=~2.76$~Te\kern-.1emV\xspace}
\newcommand{\fivenn}       {$\sqrt{s_{\mathrm{NN}}}~=~5.02$~Te\kern-.1emV\xspace}
\newcommand{\LT}           {L{\'e}vy-Tsallis\xspace}
\newcommand{\GeVc}         {Ge\kern-.1emV/$c$\xspace}
\newcommand{\MeVc}         {Me\kern-.1emV/$c$\xspace}
\newcommand{\TeV}          {Te\kern-.1emV\xspace}
\newcommand{\GeV}          {Ge\kern-.1emV\xspace}
\newcommand{\MeV}          {Me\kern-.1emV\xspace}
\newcommand{\GeVmass}      {Ge\kern-.2emV/$c^2$\xspace}
\newcommand{\MeVmass}      {Me\kern-.2emV/$c^2$\xspace}
\newcommand{\lumi}         {\ensuremath{\mathcal{L}}\xspace}

\newcommand{\ITS}          {\rm{ITS}\xspace}
\newcommand{\TOF}          {\rm{TOF}\xspace}
\newcommand{\ZDC}          {\rm{ZDC}\xspace}
\newcommand{\ZDCs}         {\rm{ZDCs}\xspace}
\newcommand{\ZNA}          {\rm{ZNA}\xspace}
\newcommand{\ZNC}          {\rm{ZNC}\xspace}
\newcommand{\SPD}          {\rm{SPD}\xspace}
\newcommand{\SDD}          {\rm{SDD}\xspace}
\newcommand{\SSD}          {\rm{SSD}\xspace}
\newcommand{\TPC}          {\rm{TPC}\xspace}
\newcommand{\TRD}          {\rm{TRD}\xspace}
\newcommand{\VZERO}        {\rm{V0}\xspace}
\newcommand{\VZEROA}       {\rm{V0A}\xspace}
\newcommand{\VZEROC}       {\rm{V0C}\xspace}
\newcommand{\Vdecay} 	   {\ensuremath{V^{0}}\xspace}

\newcommand{\ee}           {\rm \ensuremath{e^{+}e^{-}}} 
\newcommand{\pip}          {\ensuremath{\pi^{+}}\xspace}
\newcommand{\pim}          {\ensuremath{\pi^{-}}\xspace}
\newcommand{\kap}          {\ensuremath{\rm{K}^{+}}\xspace}
\newcommand{\kam}          {\ensuremath{\rm{K}^{-}}\xspace}
\newcommand{\pbar}         {\ensuremath{\rm\overline{p}}\xspace}
\newcommand{\kzero}        {\ensuremath{{\rm K}^{0}_{\rm{S}}}\xspace}
\newcommand{\lmb}          {\ensuremath{\Lambda}\xspace}
\newcommand{\almb}         {\ensuremath{\overline{\Lambda}}\xspace}
\newcommand{\Om}           {\ensuremath{\Omega^-}\xspace}
\newcommand{\Mo}           {\ensuremath{\overline{\Omega}^+}\xspace}
\newcommand{\X}            {\ensuremath{\Xi^-}\xspace}
\newcommand{\Ix}           {\ensuremath{\overline{\Xi}^+}\xspace}
\newcommand{\Xis}          {\ensuremath{\Xi^{\pm}}\xspace}
\newcommand{\Oms}          {\ensuremath{\Omega^{\pm}}\xspace}
\newcommand{\degree}       {\ensuremath{^{\rm o}}\xspace}

\newcommand{\kT}{k_{\rm T}}
\newcommand{\OmegacZero}{\rm \Omega_{c}^{0}}
\newcommand{\XicZero}{\rm \Xi_{c}^{0}}
\newcommand{\XicPlus}{\rm \Xi_{c}^{+}}
\newcommand{\XicZeroPlus}{\rm \Xi_{c}^{0,+}}
\newcommand{\Jpsi}{{\rm J}/\psi}
\newcommand{\acceff}{\rm Acc.\times\varepsilon}
\newcommand{\Dzero}{\rm D^{0}}
\newcommand{\Dplus}{\rm D^+}
\newcommand{\Dstar}{\rm D^{\ast +}}
\newcommand{\Ds}{\rm D_s^+}
\newcommand{\DtoKpi}{\rm D^0\to K^-\pi^+}
\newcommand{\DplusToKpipi}{\Dplus \to \rm K^{-} \pi^{+} \pi^{+}}
\newcommand{\DstarToDzeropi}{\Dstar \to \Dzero \pi^{+} \to K^- \pi^+\pi^{+}}
\newcommand{\DsToPhipiToKKpi}{\Ds \to \phi \pi^{+} \to K^{+}K^{-}\pi^{+}}
\newcommand{\Sigmac}{\rm \Sigma_{c}}
\newcommand{\SigmacZeroPlusPlus}{{\rm \Sigma_{c}^{0,++}}}
\newcommand{\SigmacZero}{{\rm \Sigma_{c}^{0}}}
\newcommand{\SigmacPlusPlus}{{\rm \Sigma_{c}^{++}}}
\newcommand{\SigmacZeroPlusPlusPlus}{\rm \Sigma_{c}^{0,+,++}}
\newcommand{\Lambdac}{\rm \Lambda_{c}^{+}}
\newcommand{\Bs}{\rm B_{s}^{0}}
\newcommand{\Bzero}{\rm B^{0}}
\newcommand{\Bplus}{\rm B^{+}\xspace}
\newcommand{\Hb}{\rm h_{b}\xspace}
\newcommand{\LambdacTopKpi}{\rm \Lambda_{c}^+\rightarrow pK^{-}\pi^{+}}
\newcommand{\LambdacTopKzeroS}{\rm \Lambda_{c}^{+}\rightarrow pK^{0}_{S}}
\newcommand{\pKzeroS}{\rm pK^{0}_{S}}
\newcommand{\Lambdab}{\rm \Lambda_{b}^{0}}
\newcommand{\ep}{\rm e^{-}p}
\newcommand{\pKpi}{\rm pK^{-}\pi^{+}}
\newcommand{\pKzeros}{\rm pK^{0}_{s}}
\newcommand{\ccbar}{\rm c\overline{c}}
\newcommand{\bbbar}{\mathrm{b\overline{b}}}
\newcommand{\charm}{{\rm c}}
\newcommand{\noop}[1]{}
\newcommand{\fsratio}{f_{\rm{s}}/(f_{\rm{u}} + f_{\rm{d}})}

\renewcommand{\pi}{\uppi}
\renewcommand{\phi}{\upphi}

%% file: abstract.tex
\begin{abstract}
The $\pt$-differential production cross sections of non-prompt $\Dzero$, $\Dplus$, and $\Ds$ mesons originating from beauty-hadron decays are measured in proton--proton collisions at a centre-of-mass energy $\s = 13$~TeV. The measurements are performed at midrapidity, $|y| < 0.5$, with the data sample collected by ALICE from 2016 to 2018. The results are in agreement with predictions from several perturbative QCD calculations. The fragmentation fraction of beauty quarks to strange mesons divided by the one to non-strange mesons, $\fsratio$, is found to be \sloppy\mbox{$0.114 \pm 0.016~{\rm (stat.)} \pm 0.006~{\rm (syst.)} \pm 0.003~{\rm (BR)} \pm 0.003~{\rm (extrap.)}$}. This value is compatible with previous measurements at lower centre-of-mass energies and in different collision systems in agreement with the assumption of universality of fragmentation functions. In addition, the dependence of the non-prompt D meson production on the centre-of-mass energy is investigated by comparing the results  
obtained at $\s = 5.02$ and 13~TeV, showing a hardening of the non-prompt D-meson $\pt$-differential production cross section at higher $\s$. Finally, the ${\rm b\overline{b}}$ production cross section per unit of rapidity at midrapidity is calculated from the non-prompt $\Dzero$, $\Dplus$, $\Ds$, and $\Lambdac$ hadron measurements, obtaining   ${\rm d}\sigma/{\rm d}y = 75.2\pm 3.2~(\mathrm{stat.}) \pm 5.2~(\mathrm{syst.})^{+12.3}_{-3.2} ~(\mathrm{extrap.})\text{ } \rm \upmu b \;.$

\end{abstract}

%% file: introduction.tex
\section{Introduction}
\label{sec:intro}
Measuring the production of hadrons containing heavy-flavour quarks (i.e.~charm and beauty) in proton\textendash{}proton, pp, collisions is essential to test perturbative Quantum Chromodynamics (pQCD) calculations and provide a reference for analogous measurements in heavy-ion collisions~\cite{Prino:2016cni}. The ALICE~\cite{ALICE:2012inj, ALICE:2019nxm, ALICE:2017olh, ALICE:2012gkr, ALICE:2012sxy, ALICE:2012msv, ALICE:2014aev, ALICE:2012vpz, ALICE:2018fvj, ALICE:2019nuy, ALICE:2019rmo, ALICE:2022wpn, ALICE:charm_pp13}, ATLAS~\cite{ATLAS:2015igt,ATLAS:2013cia,ATLAS:2015esn}, CMS~\cite{CMS:2021lab, PhysRevLett.131.121901, CMS:2011kew, CMS:2011pdu, CMS:2012wje}, and LHCb~\cite{LHCb:2013xam, LHCb:2015swx, LHCb:2016ikn, LHCb:2017vec, LHCb:2015qvk1, LHCb:2018jry} experiments at the LHC have measured the production of charm and beauty hadrons and their decay leptons in pp collisions at various centre-of-mass energies ($\s$) ranging from 2.76 to 13 TeV, while RHIC~\cite{PHENIX:2006tli, PHENIX:2009dpd, STAR:2012nbd, PHENIX:2017ztp}, Sp$\overline{\rm p}$S~\cite{UA1:1990vvp}, and the Tevatron~\cite{CDF:2003vmf, CDF:2004jtw, CDF:2006ipg, CDF:2009bqp} performed measurements at lower $\s$ values. The theoretical calculations rely on the factorisation of soft and hard processes~\cite{Collins:1989gx} to predict the production cross sections of charm and beauty hadrons as a function of the transverse momentum ($\pt)$ and the rapidity ($y$). According to the collinear factorisation approach, the $\pt$- and $y$-differential cross sections can be computed as the convolution of three ingredients: (i) the parton distribution functions (PDFs) describing the probability of the parton to inherit a certain fraction ($x$) of the momentum of the colliding proton; (ii) the partonic scattering cross section defining the scattering probability calculated as a perturbative series expansion in the strong coupling constant ($\alpha_{\rm s}$); (iii) the fragmentation function (FF) that describes the non-perturbative transition of a heavy-flavour quark into a hadron. The FF is parametrised from measurements performed in $\ee$ or $\ep$ collisions~\cite{Kneesch:2007ey, Kniehl:2006mw}, thus assuming the hadronisation process of charm and beauty quarks to be independent of the collision system. 

Calculations for LHC energies implementing the collinear factorisation approach, like the General-Mass Variable-Flavour-Number Scheme (GM-VFNS)~\cite{Benzke:2017yjn, Kramer:2017gct, Kniehl:2004fy, Kniehl:2005mk, Kniehl:2012ti, Helenius:2018uul} and Fixed Order plus Next-to-Leading Logarithms (FONLL)~\cite{Cacciari:1998it, Cacciari:2012ny}, provide a Next-to-Leading Order (NLO) accuracy with all-order resummation of next-to-leading logarithms. In addition, calculations with Next-to-Next-to-Leading-Order (NNLO) QCD radiative corrections are available for beauty-quark production. GM-VFNS and FONLL calculations describe within uncertainties the production of D mesons originating from charm-quark hadronisation (i.e.\ prompt) and from beauty-hadron decays (i.e.\ non-prompt) as a function of \pt at different centre-of-mass energies, as well as the measured production cross sections of heavy-flavour decay leptons and non-prompt J/$\psi$ mesons originating from beauty-hadron decays~\cite{CMS:2010nis,ALICE:2012msv,ALICE:2012inj,ALICE:2012acz, ATLAS:2013cia, CMS:2016plw, LHCb:2015foc, LHCb:2015swx, LHCb:2016ikn, LHCb:2016qpe,ALICE:2019nuy, ALICE:2023xiu}. Recent D and B meson measurements in small collision systems from the LHCb Collaboration~\cite{LHCb:2023rpm, LHCb:2022syj} indicate a larger production of strange over non-strange mesons when moving to a small to larger number of tracks produced in the collision. These results are strongly underestimated by predictions based on e$^+$e$^-$ measurements and suggest the presence of unexpected nuclear effects also in small systems. For charm and beauty baryons, however, the pQCD calculations using FF from e$^+$e$^-$ collisions severely underestimate the measured cross sections~\cite{LHCb:2011leg,ALICE:2020wfu,ALICE:2020wla,ALICE:2022ych,Acharya:2021vpo,Acharya:2021vjp,LHCb:2019fns}. For instance, the prompt $\Lambdac$-baryon production cross section at low $\pt$ and midrapidity ($|y| < 0.5$) in pp collisions at $\s = 5.02$~TeV~\cite{ALICE:2020wfu,ALICE:2020wla,CMS:2019uws, ALICE:2022ych} is underestimated by a factor of 3 to 4 by GM-VFNS calculations adopting $\Lambdac$-baryon fragmentation functions derived from the fit of OPAL data~\cite{de35328685ef47f1807976f1f15f14be}, and by a factor of 15 by the POWHEG predictions~\cite{Frixione_2007} matched with PYTHIA~6~\cite{Sjostrand:2006za} to generate the parton shower. This discrepancy challenges the assumption of universality of the hadronisation process, i.e.\ the hypothesis that the parton fragmentation is independent of the collision system and energy, and can be determined from measurements in e$^+$e$^-$ collisions. It is therefore crucial to extend the study of fragmentation fractions to different collision systems, centre-of-mass energies, and $y$ intervals.



This paper presents the $\pt$-differential production cross sections of non-prompt $\Dzero$, $\Dplus$, and $\Ds$ mesons at midrapidity in \pp collisions at $\s=13$ TeV. Sections~\ref{sec:apparatus} and~\ref{sec:analysis} are devoted to the description of the experimental apparatus and the analysis strategy employed in this study, respectively.
The sources of systematic uncertainty affecting the measurement of the production cross section and their magnitudes are detailed in Sec.~\ref{sec:systematics}.
Section~\ref{sec:results} reports the non-prompt D-meson cross sections compared to pQCD predictions, the ratios of D-meson species yields at $\sqrt{s}=13$ and 5.02~TeV, and the fragmentation fraction of beauty quarks together with the total production cross section of beauty quarks at $\s = 13$~TeV.
This latter result supersedes the one of Ref.~\cite{ALICE:NPLc13TeV} by using a non-prompt $\Dzero$-meson measurement with finer granularity in \pt\ and by including the contributions from non-prompt $\Dplus$ and $\Ds$ mesons.

%% file: apparatus.tex
\section{Experimental apparatus and data sample}
\label{sec:apparatus}
A detailed description of the ALICE detector and its performance can be found in Refs.~\cite{ALICE:2008ngc,ALICE:2014sbx}. Heavy-flavour hadron decays are reconstructed with the detectors of the central barrel, which cover the pseudorapidity range $|\eta|<0.9$ and are located inside a cylindrical solenoid that produces a magnetic field of $B = 0.5$~T along the beam direction. Charged-particle trajectories are reconstructed by the Inner Tracking System (\ITS) and the Time Projection Chamber (\TPC). The \ITS detector is composed of six layers of silicon detectors, which delivers precise measurements of track parameters near the interaction point, and additionally provides a resolution on the track impact parameter in the transverse plane better than 75~$\upmu$m for tracks with $\pt > 1$~GeV/$c$. The \TPC provides track reconstruction featuring up to 159 three-dimensional space points per track, as well as charged particle identification (PID) through specific energy loss ($\dEdx$) measurements. Additionally, the Time-Of-Flight (\TOF) detector measures the flight time of charged particles, 
providing additional constraints for the identification of the decay products of heavy-flavour hadrons. Trigger and event selections are performed using the \VZERO detector, which comprises two scintillator arrays located on either side of the collision point and that cover the pseudorapidity ranges $-3.7<\eta<-1.7$ and $2.8<\eta<5.1$.

The measurements reported in this article were performed on the data sample of \pp collisions at \linebreak \mbox{$\s=13$}~TeV collected with the ALICE experiment from 2016--2018. The \pp collisions were recorded using a minimum bias (MB) trigger that required coincident signals in both \VZERO scintillator arrays. Beam-induced background events, including beam--gas interactions and pileup of collisions from different bunch crossings, were removed offline using the timing information from the \VZERO arrays and correlating the number of measured clusters and tracks reconstructed in the two innermost layers of the ITS. Events with pileup of collisions within the same bunch crossing were eliminated by rejecting events with more than one reconstructed primary vertex. Moreover, to ensure uniform pseudorapidity acceptance, only events with a primary vertex position within $\pm 10$~cm from the nominal centre of the apparatus along the beam direction were considered. The analysed data consisted of about $1.8 \times 10^{9}$ MB collisions, corresponding to an integrated luminosity $\mathcal{L}_{\rm int}= 31.9 \pm 0.5~\text{nb}^{-1}$~\cite{ALICE-PUBLIC-2021-005}.

To correct for detector acceptance and efficiency and to train the machine learning algorithms that are used in this analysis as described later, Monte Carlo (MC) simulations of \pp collisions at the same centre-of-mass energy were utilised. These MC samples were generated using the PYTHIA~8.243~\cite{SJOSTRAND2015159} event generator, requiring the production of at least one $\ccbar$ or $\bbbar$ pair in each simulated event. The produced charm hadrons were forced to decay in the decay channels of interest. The generated particles were transported through the apparatus with GEANT3~\cite{Brun:1082634}, including a realistic description of the detector conditions during the data taking.

%% file: analysis.tex
\section{Data analysis}
\label{sec:analysis}
The D mesons and their charge conjugates were reconstructed through the following hadronic decay channels: $\Dzero\to\kam\pip$ with BR = (3.95 $\pm$ 0.03)\%, $\Dplus\to\kam\pip\pip$ with BR = (9.38 $\pm$ 0.16)\%, and $\Ds\to\phi\pip\to\kap\kam\pip$ with BR = (2.22 $\pm$ 0.06)\%~\cite{10.1093/ptep/ptac097}. The D-meson candidates were reconstructed by combining tracks with $|\eta|<0.8$ and $\pt>0.3$~GeV/$c$. 
Only tracks crossing at least 70 pad rows in the TPC, having at least one associated hit in the two innermost ITS layers, and passing the track-quality criteria described in Ref.~\cite{ALICE:2019nxm} were considered. Pions and kaons were identified by requiring the $\dEdx$ and time-of-flight signals to be compatible with the expected values within three times the detector resolution. 
The selections applied to the single tracks affect the D-meson reconstruction and acceptance as a function of the rapidity, which decreases steeply for $|y|>0.5$ at $\pt < 5$~\GeVc and $|y|>0.8$ for $\pt > 5$~\GeVc. For this reason, a $\pt$-dependent selection was applied on the D-meson rapidity to define a fiducial acceptance, $|y| < y_{\rm fid} (\pt)$, with $y_{\rm fid} (\pt)$ increasing from 0.5 to 0.8 in $0 < \pt < 5$~\GeVc and $y_{\rm fid} = 0.8$ above $5$~\GeVc. This is also taken into account when computing the acceptance term in the cross section formula.

A Boosted Decision Tree (BDT) algorithm was employed to reduce the sizeable combinatorial background and to improve the separation between the contributions of prompt and non-prompt D mesons through a multi-class classification approach~\cite{ALICE:2022pxx}. The implementation of the BDT algorithm provided by the XGBoost~\cite{10.1145/2939672.2939785, barioglio_luca_2022_7014886} library was employed for the results presented in this work.
The machine learning algorithm was provided with signal examples of D mesons from simulations based on the PYTHIA~8.243 event generator, while the background samples were obtained from the \enquote{sideband} region of the D-meson candidate invariant-mass ($M$) distributions in data. For $\Dzero$ and $\Dplus$, this region was defined as the invariant-mass intervals $ |\Delta M| > 5 \sigma$ with respect to the nominal value of their respective mass (i.e.~$M<1.80$ GeV/$c^2$ and $M>1.95$ GeV/$c^2$ for $\Dzero$ and $M<1.82$ GeV/$c^2$ and $M>1.92$ GeV/$c^2$ for $\Dplus$). 
For $\Ds$ mesons, the regions $M < 1.82$ GeV/$c^2$ and $M > 2.01$~GeV/$c^2$ were considered, corresponding to masses of the order of $5\sigma$ below the nominal $\Dplus$ mass and $5\sigma$ above the nominal $\Ds$ mass. This selection was tuned to efficiently reject the contribution from $\Dplus$ mesons decaying in the same decay channel. Loose selection criteria were applied to the D-meson candidates before the BDT training, following the same procedures described in Refs.~\cite{ALICE:2019nxm, ALICE:2021mgk,  ALICE:2022xrg}. For the $\Ds$-meson candidates, an additional selection was applied on the reconstructed value of the $\mathrm{K^{+}K^{-}}$ invariant mass to be consistent with the Particle Data Group world average of the $\phi$-meson rest mass ($M(\phi) = 1019.461 \pm 0.016$~\cite{10.1093/ptep/ptac097}) with $\pm 15$~MeV/$c^2$. The variables provided to the BDT algorithm to classify the candidates as either D mesons originating from beauty-hadron decays, prompt D mesons, or background candidates, include those related to the decay-vertex topology and single-track PID, which have proven to strongly influence the score assigned by the BDT. In particular, the impact of each input variable was assessed via the SHAP package~\cite{NIPS2017_7062}. The complete list of input variables can be found in Ref.~\cite{ALICE:2021mgk}. The information about variables related to the decay-vertex topology was found to be the most relevant to discriminate signal and background candidates, as expected.
Independent BDTs were trained for the different $\pt$ intervals of the analyses and the resulting optimised algorithms were applied to the data, where the type of candidate is unknown. The BDT provides three different outputs, one for each class, representing the estimated probability of a candidate to be a 
prompt D meson, a non-prompt D meson, or combinatorial background. Selections were applied on both the probability to be a non-prompt D meson and on the probability to be combinatorial background. The first selection was meant to enhance the non-prompt contribution within the selected signal, and the second one was employed to reject the largest possible amount of combinatorial background as possible, while preserving the signal candidates. Different selection criteria on the BDT outputs were used in the analysis in order to build samples with different fractions of non-prompt (and prompt) D mesons~\cite{ALICE:2021mgk, ALICE:2022pxx, ALICE:2022xrg}. 

\begin{figure}[t!]
    \centering
    \includegraphics[width=0.49\textwidth]{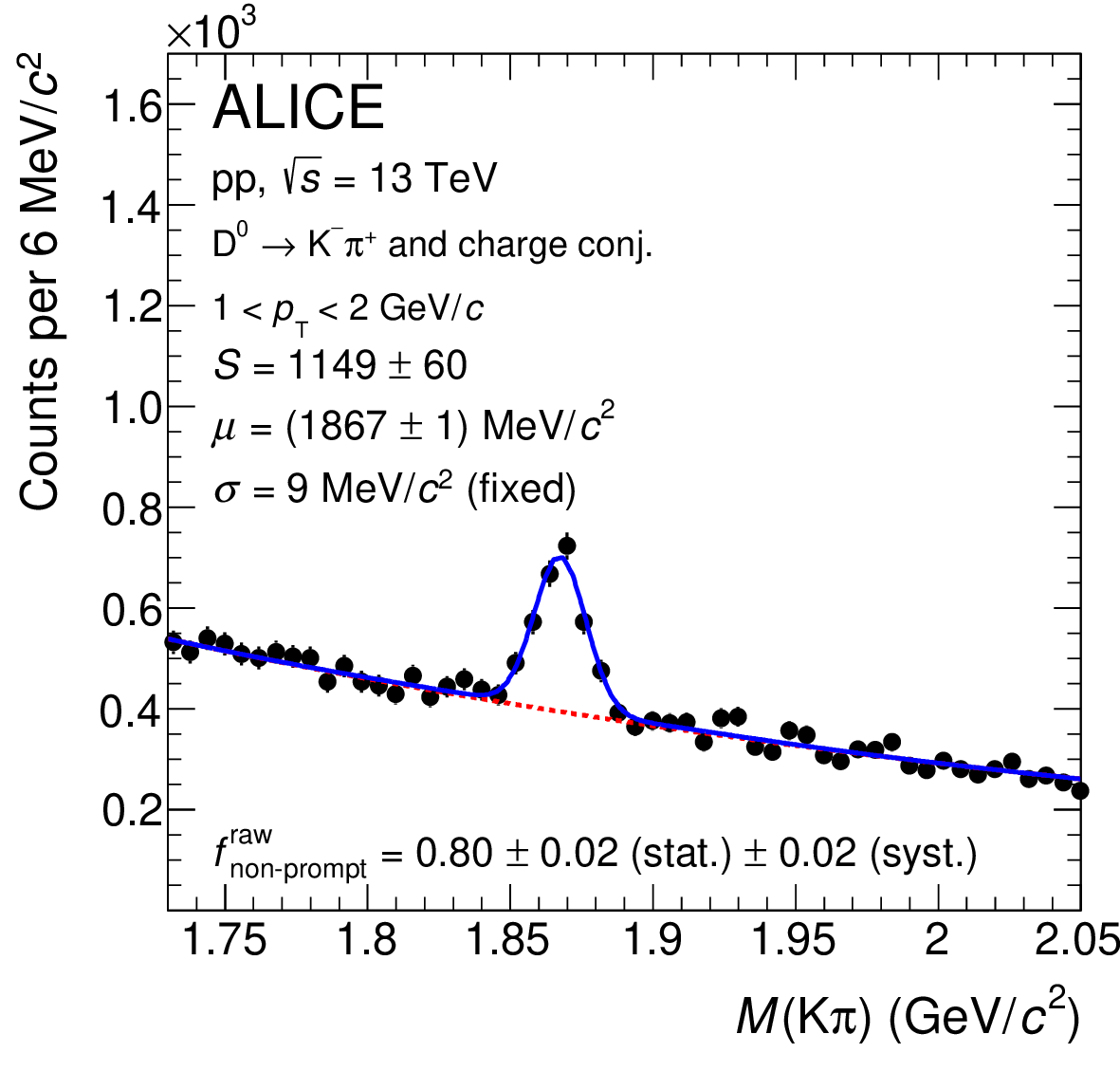}
    \includegraphics[width=0.49\textwidth]{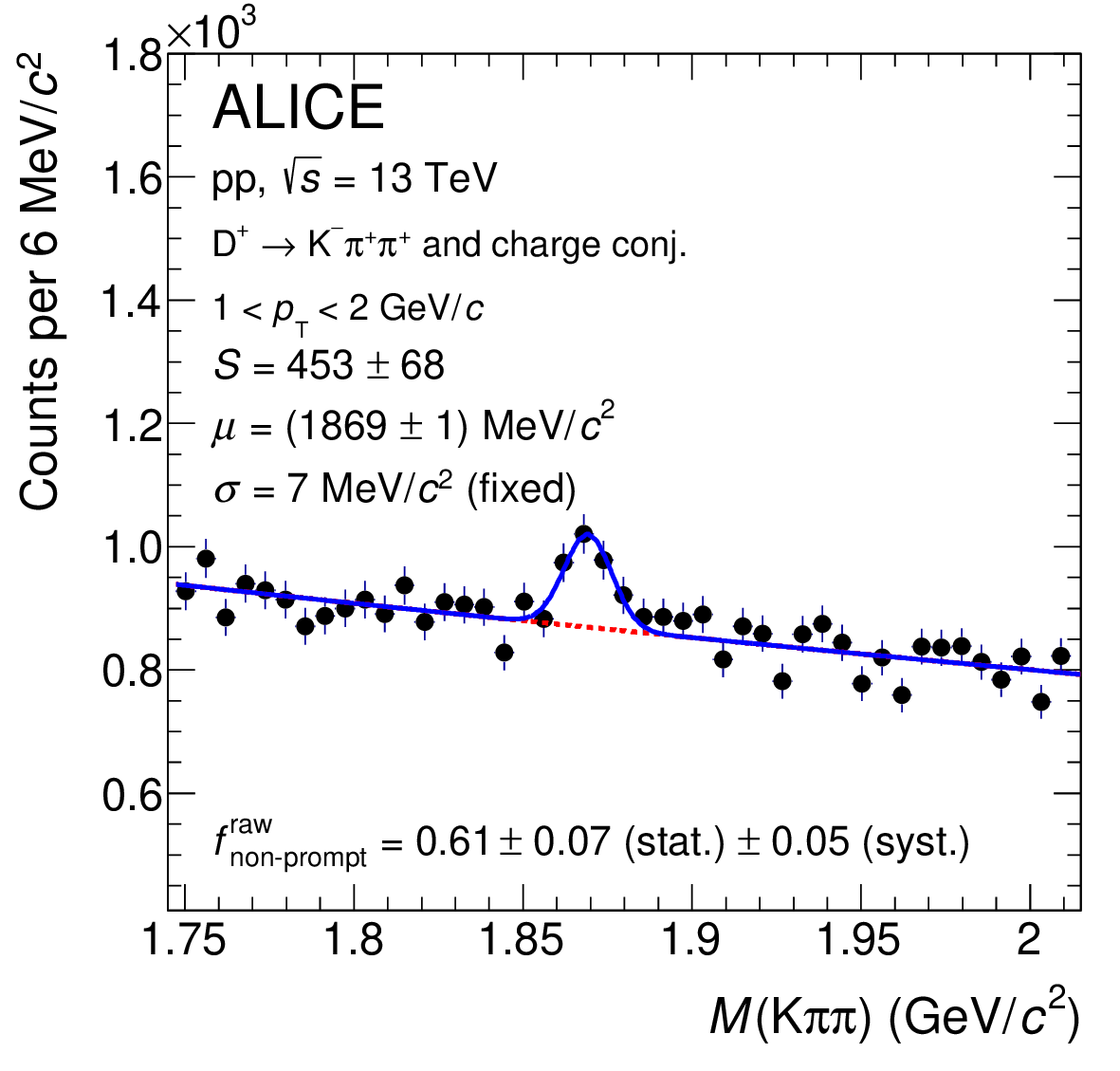}
    \includegraphics[width=0.49\textwidth]{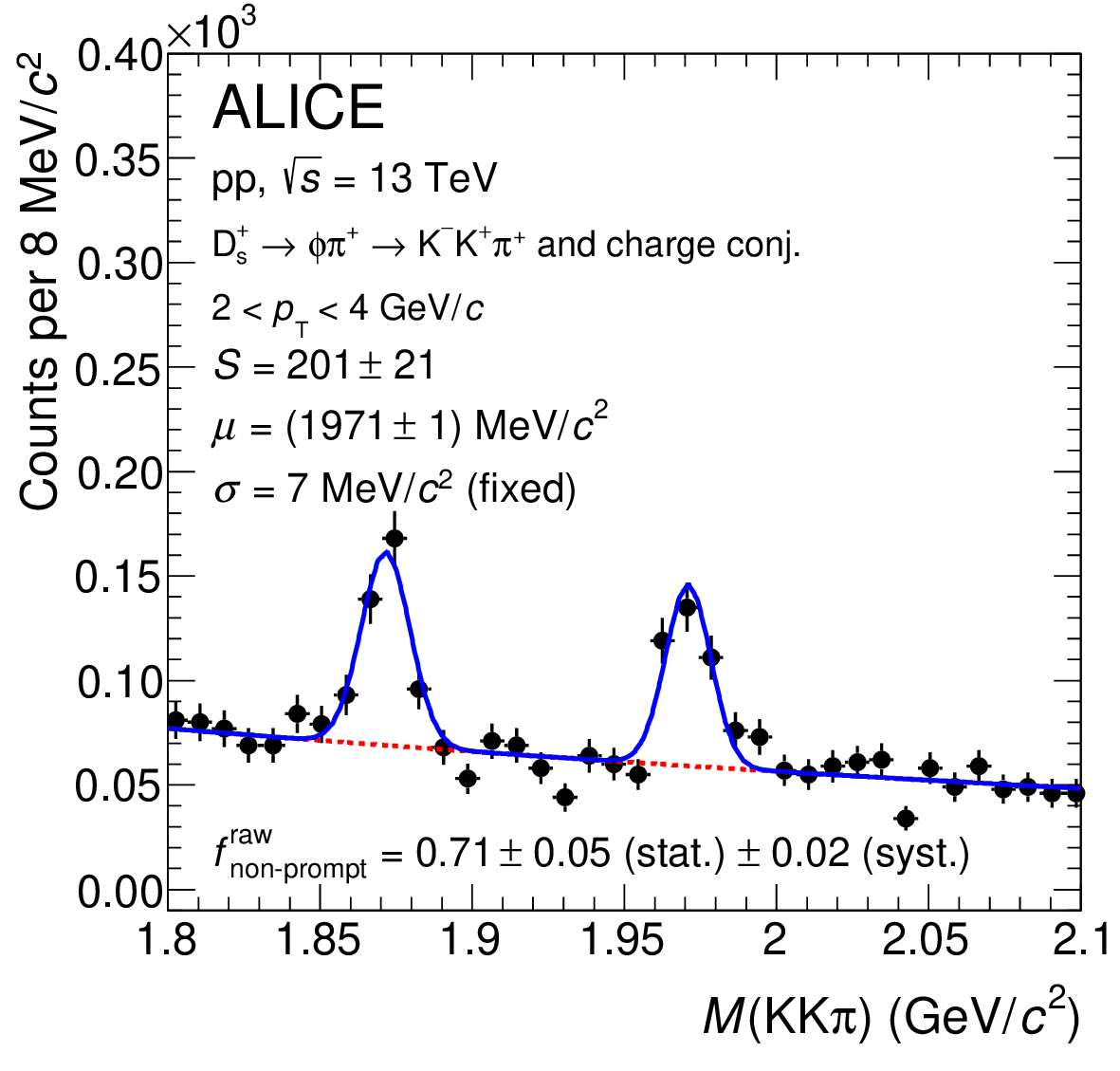}
    \caption{Invariant-mass distributions of non-prompt enhanced $\Dzero$- (top-left), $\Dplus$- (top-right), and $\Ds$-meson (bottom) candidates, and their charge conjugates in $1< \pt < 2$ GeV/$c$, $1< \pt < 2$ GeV/$c$, and $2< \pt < 4$ GeV/$c$, respectively. For the $\Ds$ meson, the left-side peak emerges due to the contribution of the $\Dplus$ meson decaying in the same channel as the $\Ds$ meson.
    The blue solid line shows the total fit function and the red dashed line the combinatorial-background contribution. The values of the mean ($\mu$), width ($\sigma$), and raw yield ($S$) of the signal peak are reported together with their statistical uncertainties resulting from the fit. The width is fixed to the one obtained from the prompt-enhanced sample.
    The fraction of non-prompt candidates in the measured raw yield is reported with its statistical and systematic uncertainties.}
    \label{fig:minv_mesons}
\end{figure}

The raw yields of D mesons (sum of particles and antiparticles) were measured in the \linebreak range \mbox{$1 < \pt < 24$}~GeV/$c$ for $\Dzero$ and $\Dplus$ mesons and \mbox{$2 < \pt < 24$~GeV/$c$} for $\Ds$ mesons. The raw yields were extracted by performing a binned maximum-likelihood fit of the invariant mass distributions of candidates satisfying the BDT selection criteria. These criteria enhance the fraction of non-prompt D mesons, $f_{\rm non\text{-}prompt}^{\rm raw}$, in the sample, referred to as the non-prompt enhanced sample.
The signal peak was parameterised with a Gaussian function, whose width was constrained to match the one extracted from fits to the prompt-enhanced sample evaluated in this work, i.e.~candidates passing BDT selection criteria providing a significant fraction of prompt D mesons. For the $\Dzero$ meson, the contribution of signal candidates to the invariant-mass distribution with the wrong mass assigned to the $\Dzero$-decay tracks (reflections) was included in the fit. It was estimated based on the invariant-mass distributions of the reflected signal in the simulation, which were described as the sum of two Gaussian functions. The contribution of reflections to the raw yield is about 0.5--1.5$\%$, depending on $\pt$. 
This constraint improved the stability of the fits because the prompt-enhanced sample has a higher statistical significance compared to the non-prompt enhanced one due to the larger abundance of prompt D mesons compared to non-prompt D mesons.
The background was modelled with an exponential function. In the fit to the invariant-mass distributions of $\Ds$-meson candidates, an independent Gaussian function was adopted to model the peak related to $\Dplus\to \phi\pip \to \kap\kam\pip$ decays. Figure~\ref{fig:minv_mesons} shows examples of fits to the invariant-mass distributions of $\Dzero$, $\Dplus$, and $\Ds$ candidates for the non-prompt D-meson enhanced samples in the lowest $\pt$ interval accessible in the respective analyses.

The $\pt$-differential cross section of non-prompt charm hadrons at midrapidity was computed as:

\begin{equation}
    \left. \frac{{\rm d}\sigma^{\rm D}_{\rm non\mbox{-}prompt}}{{\rm d}\pt} \right|_{|y|<0.5} = \frac12\frac{1}{\Delta\pt} \frac{f_{\rm non\mbox{-}prompt}^{\rm raw} N^{\rm D+\overline{D}}_{|y|<y_{\rm fid}}}{c_{\Delta y}({\rm Acc}\times \varepsilon)_{\rm non\mbox{-}prompt}}\cdot\frac{1}{\rm BR}\cdot\frac{1}{\mathcal{L}_{\rm int}}\;.
    \label{eq:cross_section}
\end{equation}

The term $N^{\rm D+\overline{D}}_{|y|<y_{\rm fid}}$ refers to the raw yields in the various \pt\ intervals, extracted as described above. This quantity was then scaled by the non-prompt fraction $f_{\rm non\text{-}prompt}^{\rm raw}$ to account for prompt D-meson signals in the raw yield, and divided by 2 to obtain the averaged yields between particles and antiparticles. The raw yield was corrected by the $c_{\Delta y}({\rm Acc}\times \varepsilon)_{\rm non\text{-}prompt}$ term, which accounts for the fiducial interval in rapidity ($c_{\Delta y}\simeq 2\mathrm{y_{fid}}$), the detector acceptance, and the reconstruction and selection efficiency of the non-prompt D-meson signal. The production cross section in each $\pt$ interval was then obtained by scaling the corrected yield of non-prompt D mesons by the $\pt$-interval width ($\Delta\pt$), the branching ratio of the decay channel chosen to reconstruct the signal (BR), and the integrated luminosity ($\mathcal{L}_{\rm int}$).

\begin{figure}[t!]
    \centering
    \includegraphics[width=0.49\textwidth]{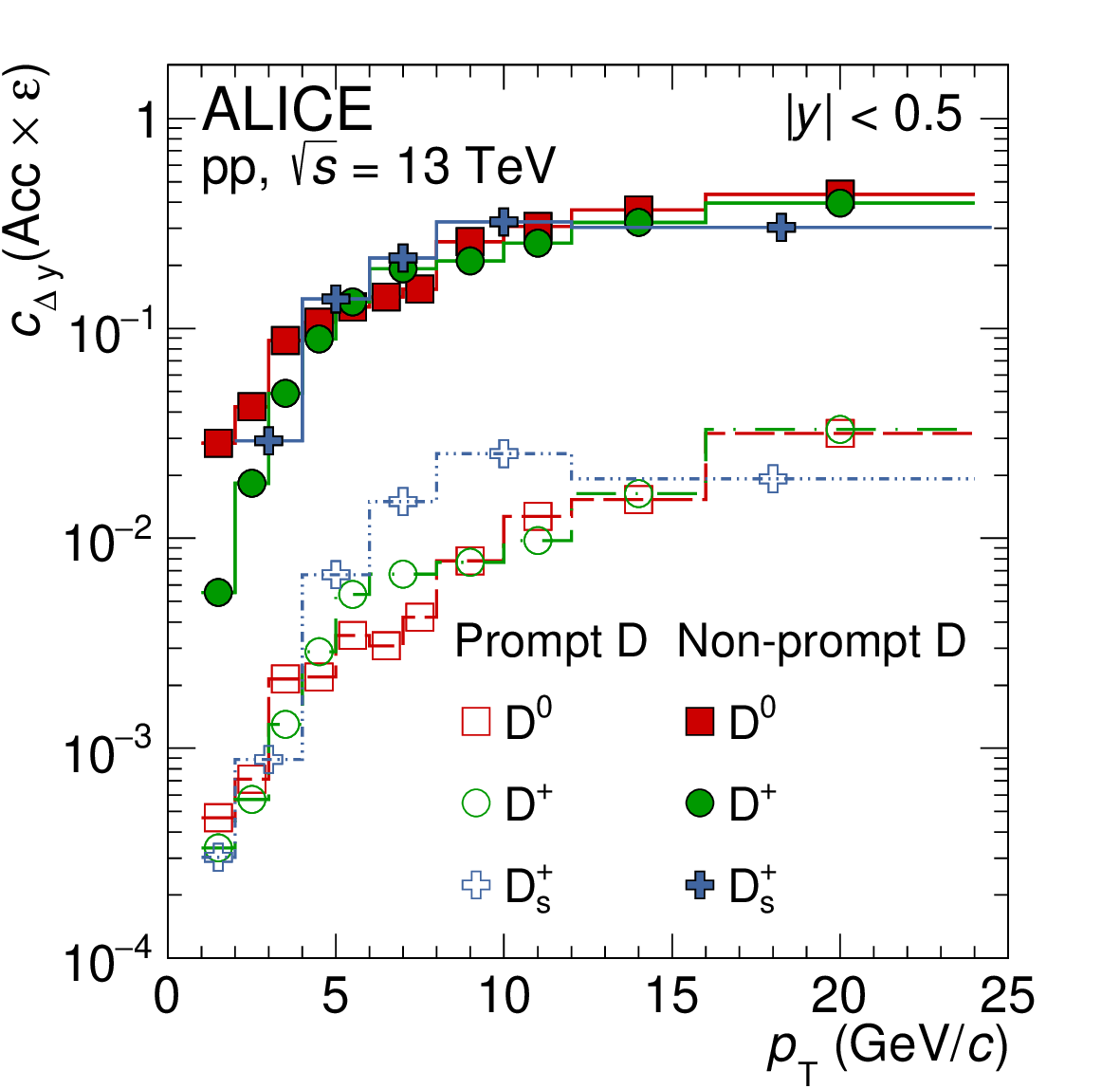}
    \includegraphics[width=0.49\textwidth]{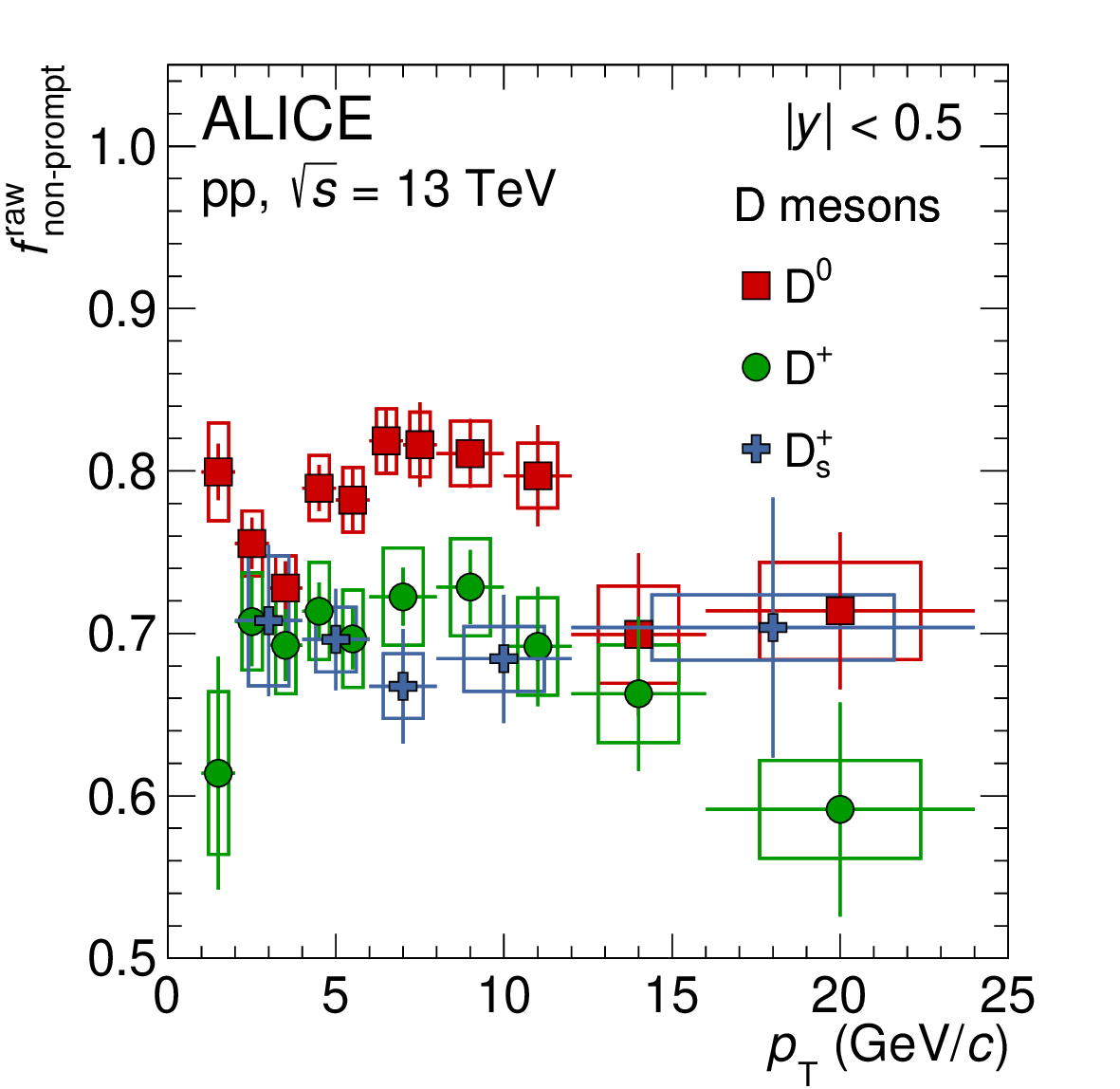}
    \caption{Left: $c_{\Delta y}({\rm Acc}\times \varepsilon)$ factors of the selection criteria of the analyses for prompt and non-prompt D mesons as a function of $\pt$. Right: raw fraction of non-prompt $\Dzero$, $\Dplus$, and $\Ds$ mesons in the raw yield as a function of $\pt$. The vertical bars and boxes display the statistical and systematic uncertainties, respectively.}
    \label{fig:npfFD_mesons}
\end{figure}

The $c_{\Delta y}({\rm Acc}\times \varepsilon)_{\rm non\text{-}prompt}$ and the $f_{\rm non\text{-}prompt}^{\rm raw}$ terms for $\Dzero$, $\Dplus$, and $\Ds$ mesons are reported as a function of $\pt$ in the left and right panels of Fig.~\ref{fig:npfFD_mesons}, respectively.
The $({\rm Acc}\times \varepsilon)_{\rm non\text{-}prompt}$ is obtained as the product of the selection efficiency $\epsilon$, accounting for the D mesons simulated with PYTHIA~8 surviving the selection criteria, and the geometrical acceptance of the experimental apparatus estimated with GEANT3. The $c_{\Delta y}$ term is introduced to normalise the D-meson yield to one unit of rapidity, thus accounting for the rapidity coverage of the measurements in $|y| < \mathrm{y_{fid}}(\pt)$, and is found to be compatible with unity in PYTHIA simulations. The fraction $f_{\rm non\text{-}prompt}^{\rm raw}$ was estimated with a data-driven procedure based on constructing data sub-samples with different abundances of prompt and non-prompt candidates. The samples were built by gradually varying solely the selection criterion on the BDT output related to the candidate's estimated probability to be a non-prompt D meson, while keeping the criterion on the probability to be background fixed to the nominal one. An equation relates the BDT selection efficiency of prompt and non-prompt D mesons and the extracted raw yields from each BDT selection criterion to the true yields of prompt and non-prompt D mesons, forming a system of equations used to estimate the $f_{\rm non\text{-}prompt}^{\rm raw}$. The effectiveness of this method was demonstrated in previous similar analyses, as documented in Refs.~\cite{ALICE:2021mgk, ALICE:2022pxx, ALICE:2022xrg}.
Fractions of non-prompt candidates larger than 70\% (60\%) were obtained for $\Dzero$ ($\Dplus$ and $\Ds$) mesons in all the $\pt$ intervals of the analysis. This shows that the BDT-based selections substantially enhance the non-prompt component in the raw yields with respect to the naturally produced fractions of non-prompt mesons, which range from 5\% at low $\pt$ to 15\% at high $\pt$.

%% file: systematics.tex
\section{Systematic uncertainties}
\label{sec:systematics}
The following sources of systematic uncertainty were considered for the non-prompt D-meson production cross sections: (i) the raw-yield extraction, (ii) track-reconstruction efficiency estimation, (iii) PID efficiency evaluation, (iv) non-prompt fraction estimation, (v) D-meson selection efficiency determination, and (vi) D-meson simulated $\pt$ shape estimation.
The sources (ii)--(vi) originate from possible differences between data and simulation due to imperfections in modelling particle interactions, the description of the detector response and alignment, or the underlying physics processes in the simulation. The resulting systematic uncertainties on the non-prompt D-meson production cross section in representative $\pt$ intervals are summarised in Table~\ref{tab:syst_Dmesons}. The systematic uncertainty assigned to the measured \pt-differential cross section is computed as the sum in quadrature of the uncertainties listed below.

The systematic uncertainty on the raw-yield extraction was assessed by varying the background fit function (linear and parabolic), the binning of the invariant-mass distribution, and the upper and lower fit limits. The sensitivity to the line shape of the D-meson peak was tested by comparing the raw-yield values from the fits with those obtained by counting the candidates in the invariant-mass region of the signal after subtracting the background estimated from a fit to the sidebands. For non-prompt $\Dzero$ mesons, an additional contribution due to the description of signal reflections in the invariant-mass distribution was estimated by varying the shape and the normalisation of the templates used for the reflections in the invariant-mass fits. The magnitude of this source of systematic uncertainty is assessed by considering the shift and the root mean square of the trial raw yield distribution with respect to the reference value and it ranges between 2\% and 10\% depending on the D-meson species and \pt.

The systematic uncertainty on the track-prolongation reconstruction efficiency 
accounts for possible discrepancies between data and MC in the TPC--ITS prolongation efficiency and the selection efficiency on track-quality criteria in the TPC. The per-track systematic uncertainties were estimated by varying the track-quality selection criteria and comparing the TPC tracks' prolongation probabilities to the ITS clusters in data and simulations. They were then propagated to the non-prompt D meson systematic uncertainty via their decay kinematics. Furthermore, a non-optimal description of the material budget in MC 
could cause a bias in the estimation of the selection efficiency. This discrepancy was handled by comparing the selection efficiency obtained using simulations with different material budgets. 
The magnitude of this source of systematic uncertainty depends on $\pt$, and it ranges from 4\% to 6\% for the two-body decays of $\Dzero$ mesons and from 5\% to 8\% for $\Dplus$ and $\Ds$ mesons, which are reconstructed via three-particle decay channels.

A possible systematic uncertainty on the PID selection efficiency 
was also considered. This source was evaluated in the prompt D-meson analysis~\cite{ALICE:charm_pp13}, and it was found to be negligible for the adopted PID strategy for all the three D meson species.

The systematic uncertainties on the estimation of the BDT-selection efficiency 
and the non-prompt D-meson fraction
account for discrepancies between data and MC simulations in the distributions of the variables used in the BDT-model training (i.e.~the D-meson decay-vertex topology, kinematics, and PID variables). The former was studied by repeating the entire analysis, varying the BDT selection criteria. The uncertainty was computed as the quadratic sum of the root mean square (RMS) and the shift in the distribution of the corrected yields derived from the variation of the BDT selection in relation to the reference value. The systematic uncertainty of the non-prompt fraction was estimated by varying the configuration of the BDT selections included in the data-driven method described in Section~\ref{sec:analysis}. The selections based on the BDT probabilities were varied considering looser and tighter conditions on the probability of the D-meson candidates being non-prompt D mesons. The non-prompt D-meson fraction was computed for each configuration, and the systematic uncertainty was assigned considering the variation with respect to the reference case. The magnitude of these uncertainties ranges between 2 and 5\% for the non-prompt D-meson fraction and from 2 to 10\% for the BDT selection efficiency, depending on the D-meson species and $\pt$. In particular, the uncertainty due to the selection efficiency is larger at low $\pt$ where more stringent selection criteria are applied.

 The calculation of the $({\rm Acc}\times \varepsilon)$ factor can be influenced by differences between the hadron $\pt$ distributions generated in the simulation and those in data. To estimate the uncertainty the simulated $\pt$ distributions were weighted to match the $\pt$ spectra derived from FONLL calculations instead of those from PYTHIA 8 simulations. In particular, the FONLL and PYTHIA~8.243 predictions for D mesons were used to compute the weights for prompt D mesons, while for non-prompt D mesons, the spectra of the beauty-hadron parents were considered. The $({\rm Acc}\times \varepsilon)$ correction factor and the raw non-prompt D meson fraction were then recomputed employing the weighted spectra, and a comparison was made between the reference and the weighted cases. The difference between these two results was assigned as a systematic uncertainty. The magnitude of this uncertainty ranges between 1 and 7\% and it is larger in the low-$\pt$ intervals where the efficiency steeply increases with $\pt$.

Finally, the uncertainties on the BR of the analysed decay channel of the three D-meson species (0.8\% for $\Dzero$, 1.7\% for $\Dplus$, and 2.7\% for $\Ds$)~\cite{10.1093/ptep/ptac097} and the integrated luminosity in pp collisions (1.6\%)~\cite{ALICE-PUBLIC-2021-005} were considered.

\begin{table}[tb!]
    \centering
    \caption{Relative systematic uncertainties of the measured cross sections of non-prompt $\Dzero$, $\Dplus$, and $\Ds$ mesons at midrapidity in \pp collisions at $\s=13$ TeV. The values reported in the table refer to representative $\pt$ intervals of the different channels.}
    \begin{tabular}{c c c | c c | c c}
        \toprule
         & \multicolumn{2}{c|}{$\Dzero$} &\multicolumn{2}{c|}{$\Dplus$} & \multicolumn{2}{c}{$\Ds$} \tabularnewline

         $\pt$ (GeV/$c$) &
         1--2 & 12--16 & 
         1--2 & 12--16 & 
         2--4   & 12--24 \tabularnewline 

        \midrule
         Signal extraction &
         4\% & 3\% & 
         5\% & 5\% & 
         3\% & 5\% \tabularnewline 

         \hline
         Tracking efficiency &
         4\% & 5\% & 
         6\% & 8\% & 
         5\% & 8\% \tabularnewline 

         \hline
         Non-prompt fraction& 
         2\% & 2\% & 
         5\% & 3\% & 
         4\% & 2\% \tabularnewline 

         \hline
         Selection efficiency &
         4\% & 2\% & 
         10\% & 4\% & 
         3\% & 3\% \tabularnewline 

         \hline
         MC $\pt$ shape &
         6\% & 1\% & 
         7\% & 2\% & 
         3\% & 2\% \tabularnewline 
         \midrule

         Branching ratio &
         \multicolumn{2}{c|}{0.8\%} & 
         \multicolumn{2}{c|}{1.7\%} & 
         \multicolumn{2}{c}{2.7\%} \tabularnewline 

         \hline
         Luminosity &
         \multicolumn{6}{c}{1.6\%} \tabularnewline
         \bottomrule
         Total uncertainty &
         $9.5\%$ & $7.9\%$ & 
         $15.5\%$ & $12.0\%$ & 
         $8.7\%$ & $10.6\%$ 
         \tabularnewline
         \bottomrule
    \end{tabular}
    \label{tab:syst_Dmesons}
\end{table}

%% file: results.tex
\section{Results}
\label{sec:results}

\input{results_DmesonCrossSection.tex}

\input{results_DmesonRatios.tex}

\input{results_FFbbbar.tex}

%% file: results_DmesonCrossSection.tex
\subsection{Non-prompt D-meson $\mathbf{\textit{p}_{\rm T}}$-differential cross sections}
\label{sec:DmesonCrossSection}

\begin{figure}[tb!]
    \centering
    \includegraphics[width=0.9\textwidth]{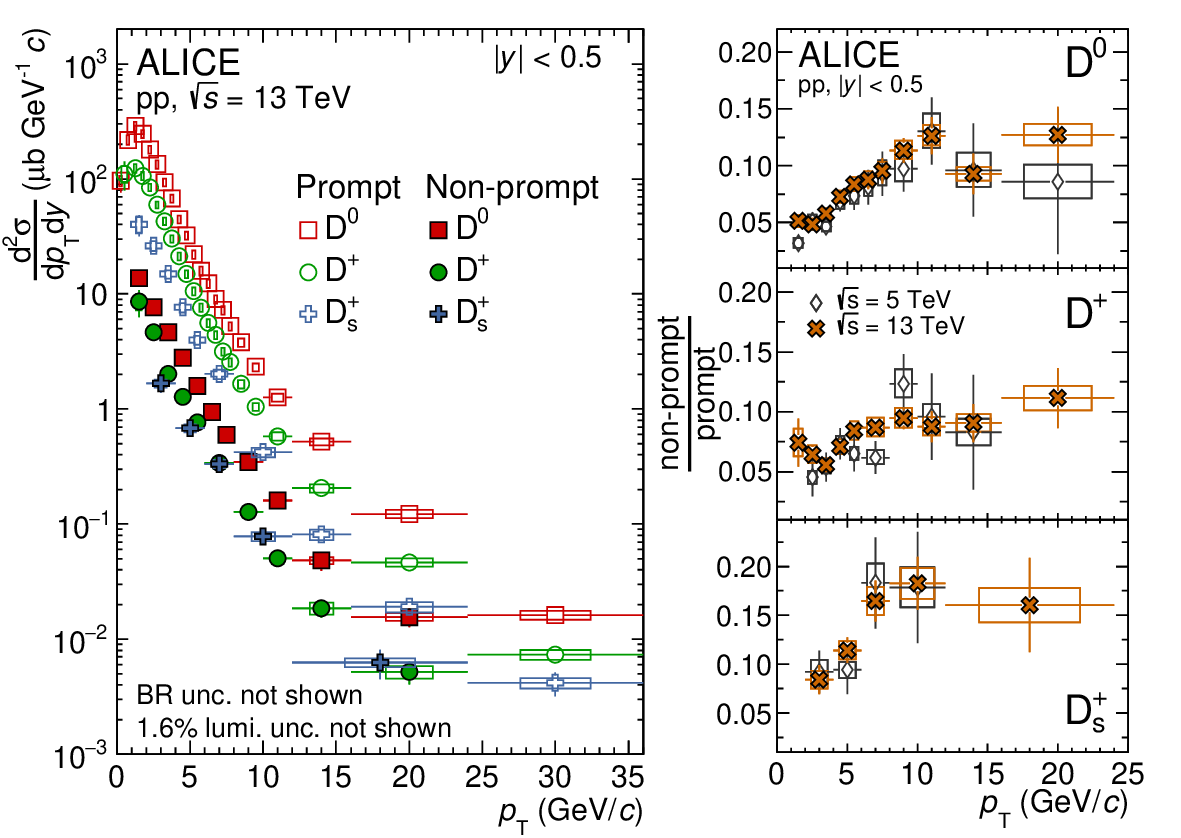}
    \caption{Left: $\pt$-differential production cross sections of prompt~\cite{ALICE:charm_pp13} and non-prompt $\Dzero$, $\Dplus$, and $\Ds$ mesons at midrapidity, $|y|<0.5$, in pp collisions at $\s=13$ TeV. Right: ratios between the production cross sections of non-prompt and prompt $\Dzero$ (top), $\Dplus$ (middle), and $\Ds$ (bottom) mesons at $\s = 5.02$~\cite{ALICE:2021mgk} and $13$~TeV. The vertical bars and empty boxes represent the statistical and systematic uncertainties, respectively.}
    \label{fig:mesonCrossSection}
\end{figure}

The $\pt$-differential production cross sections of non-prompt and prompt $\Dzero$, $\Dplus$, and $\Ds$ mesons at midrapidity, $|y|<0.5$, in pp collisions at a centre-of-mass energy of 13 TeV are shown in the left panel of Fig.~\ref{fig:mesonCrossSection}. The prompt D-meson cross sections are from Ref.~\cite{ALICE:charm_pp13}. The vertical lines and empty boxes represent the statistical and systematic uncertainties. Note that the systematic uncertainties shown in the plots do not include contributions from luminosity and branching ratio.

The right panels of Fig.~\ref{fig:mesonCrossSection} display the ratios among the $\pt$-differential production cross sections of non-prompt and prompt $\Dzero$ (top panel), $\Dplus$ (middle panel), and $\Ds$ (bottom panel) mesons at both \mbox{$\s = 5.02$}~\cite{ALICE:2021mgk} and 13~TeV. The systematic uncertainties of the ratio calculations are treated as uncorrelated among prompt and non-prompt D mesons, except for those associated with the tracking, luminosity, and BR, which are considered as fully correlated.

The uncertainties on the luminosity and the BR cancel out in the ratio. The production of prompt D mesons exceeds that of non-prompt D mesons for all D-meson species by about a factor of 20 at low $\pt$. The larger abundance of prompt charm mesons is expected owing to the lower mass of charm with respect to beauty quarks. The ratios exhibit an increasing trend as a function of $\pt$ up to approximately 12~GeV/$c$, consistent within uncertainties between the two centre-of-mass energies. This rise reflects the harder transverse-momentum spectrum of beauty hadrons (h$_{\rm b}$) with respect to that of prompt charm mesons.

\begin{figure}[h!]
    \centering
    \includegraphics[width=0.44\textwidth]{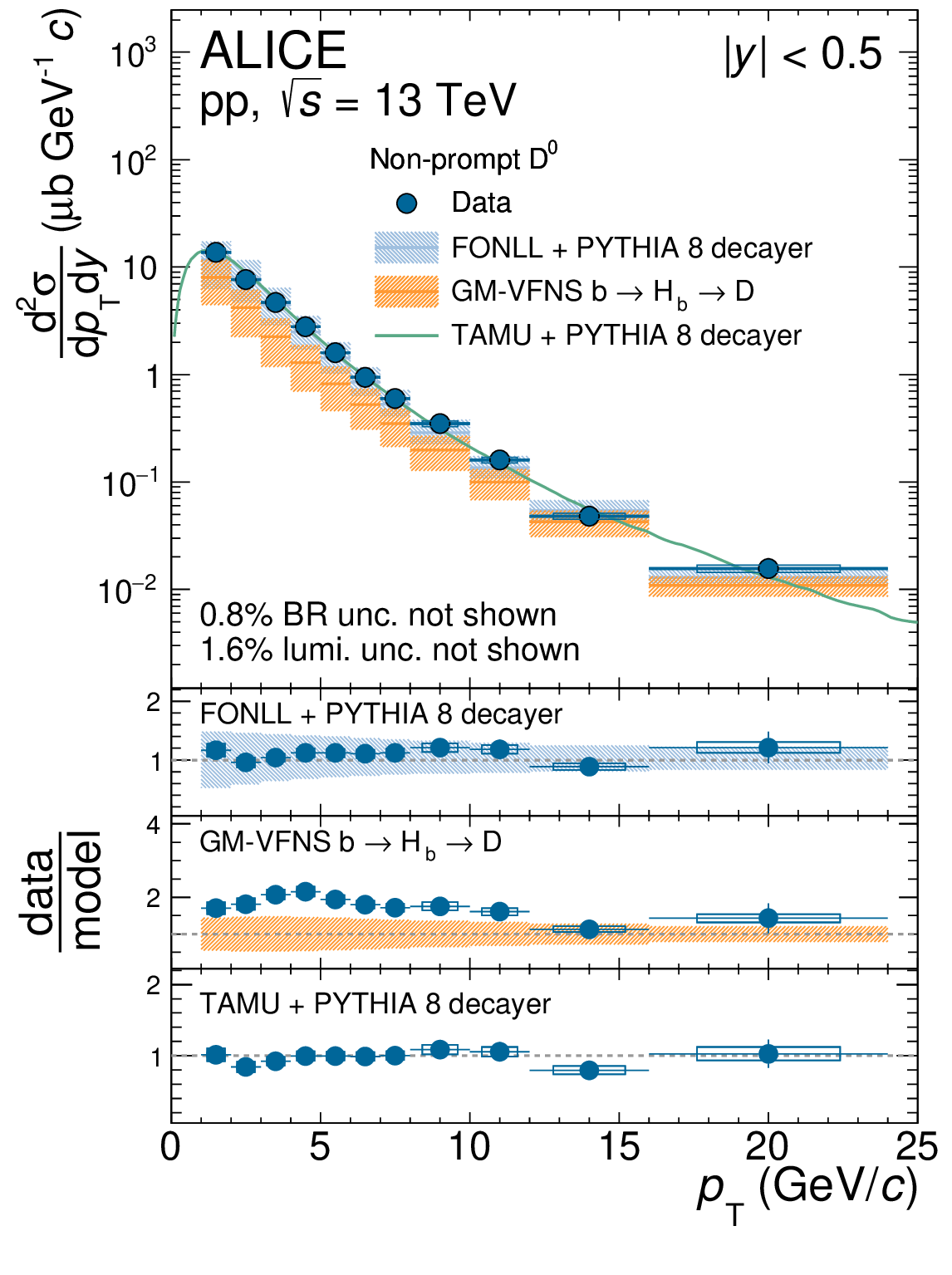}
    \includegraphics[width=0.44\textwidth]{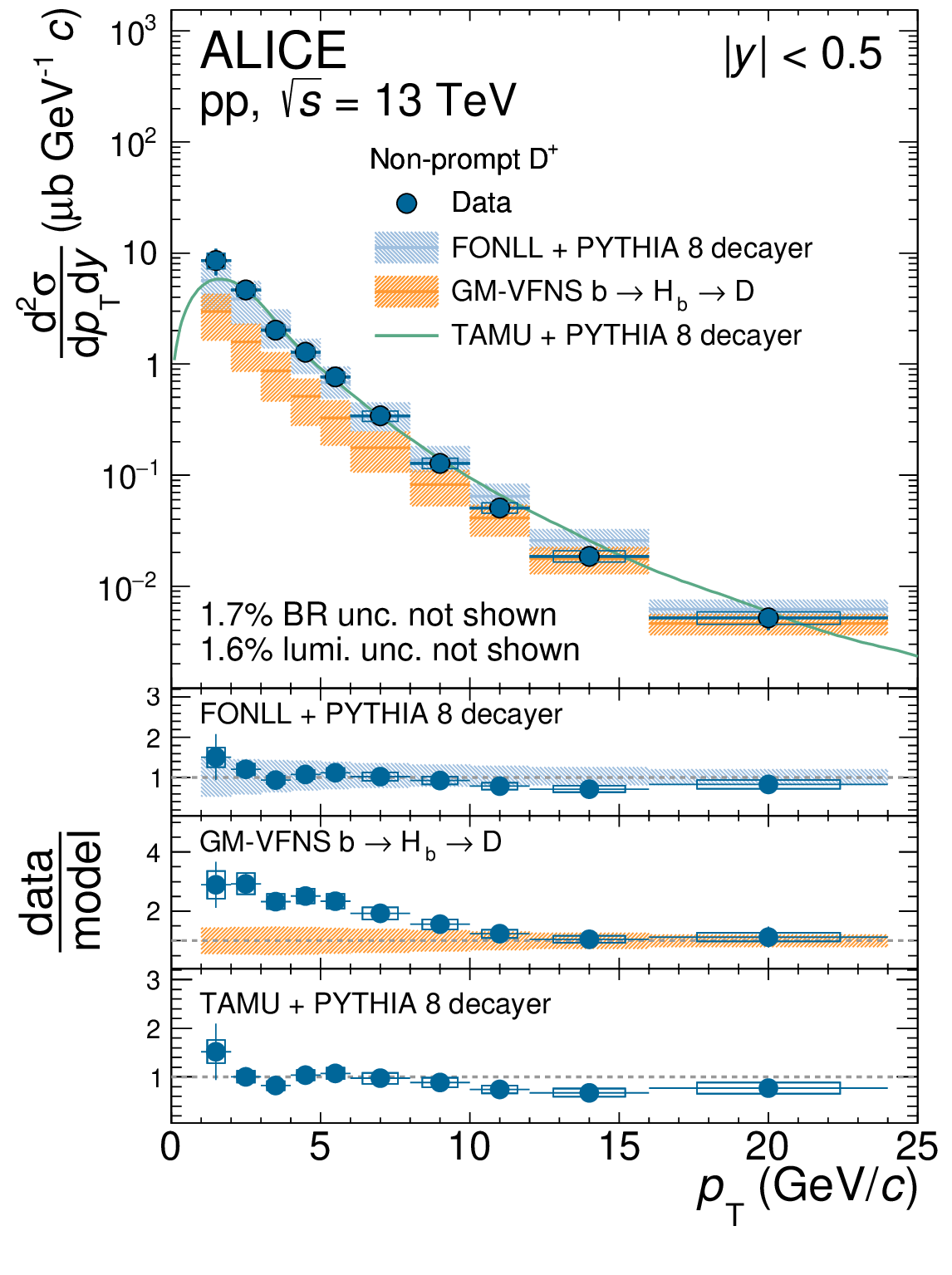}
    \includegraphics[width=0.44\textwidth]{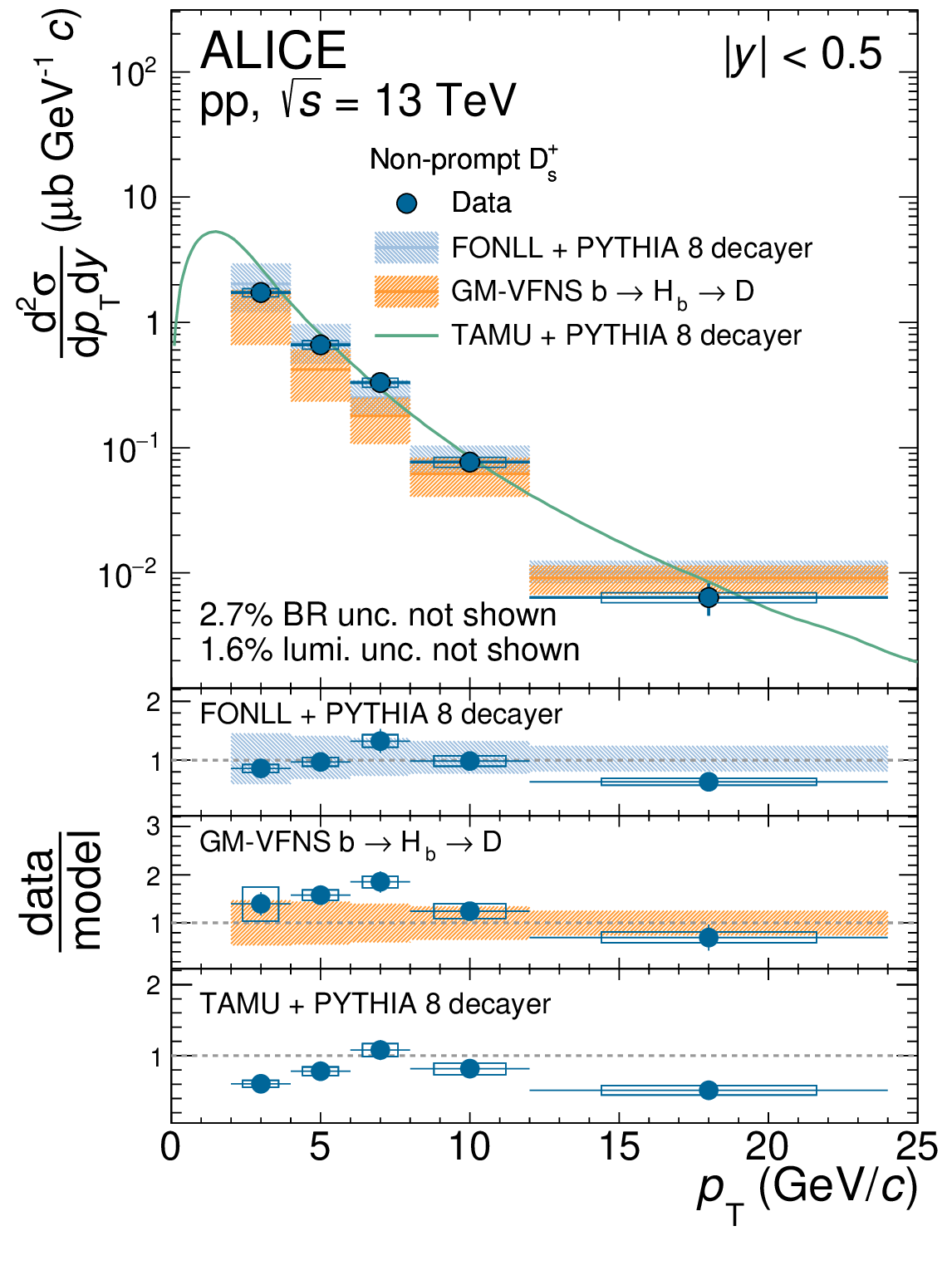}
    \caption{$\pt$-differential production cross sections of non-prompt $\Dzero$ (top-left), $\Dplus$ (top-right), and $\Ds$ (bottom) mesons at midrapidity, $|y|<0.5$, in pp collisions at $\s=13$ TeV compared with the predictions from FONLL + PYTHIA~8~\cite{Cacciari:1998it, Cacciari:2012ny}, GM-VFNS~\cite{Benzke:2017yjn, Kramer:2017gct}, and TAMU + PYTHIA~8~\cite{He:2022tod} calculations. The lower panels report the data-to-model ratios. The vertical bars and empty boxes represent the statistical and systematic uncertainties, respectively.}
    \label{fig:mesonCrossSection_vs_models}
\end{figure}

In Fig.~\ref{fig:mesonCrossSection_vs_models}, the $\pt$-differential cross sections of non-prompt D mesons are compared with the predictions obtained from FONLL~\cite{Cacciari:1998it, Cacciari:2012ny} and GM-VFNS~\cite{Benzke:2017yjn, Kramer:2017gct, Bolzoni:2013vya} calculations. In these models, the fragmentation fractions of beauty quarks into different beauty--hadron species, denoted as $f(\rm b \rightarrow B)$, are derived from $\ee$ collisions~\cite{Kneesch:2007ey}. The results of calculations based on the TAMU model~\cite{He:2022tod},  which adopts the $\pt$-differential beauty-quark cross section from FONLL along with the same fragmentation functions employed in FONLL and a statistical hadronisation approach for $f(\rm b \rightarrow B)$, are also shown in Fig.~\ref{fig:mesonCrossSection_vs_models}.
In this statistical approach, chemical equilibrium among beauty--hadron species is assumed, and the relative abundances of different hadrons are determined by their masses and a universal hadronisation temperature. In the case of FONLL and TAMU predictions, the resulting beauty hadron cross section is convoluted with the decay kinematics and branching ratios of $\rm h_{b} \rightarrow h_{c} + X$ obtained using PYTHIA 8.243~\cite{SJOSTRAND2015159, Skands:2014pea} to compute the non-prompt D-meson cross section. In the context of GM-VFNS, the transition from the beauty quark to the observed charm-hadron state is described via a two-step approach. This involves the hadronisation of the beauty quark into the beauty hadron ($\rm{b} \rightarrow \rm h_{\rm b}$) and its subsequent decay into the measured charm meson ($\rm h_{\rm b} \rightarrow \rm D + X$). Even though an alternative decay path that accomplishes this transition in a single step exists ($\rm{b}\rightarrow \rm D + X$), it significantly underestimates the predictions derived from the former approach and the measured cross sections, as already observed in Ref.~\cite{ALICE:2021mgk}.
The theoretical uncertainties associated with the FONLL and GM-VFNS predictions are shown as boxes. These uncertainties encompass variations in the factorisation scale ($\mu_{\mathrm{F}}$), renormalisation scale ($\mu_{\mathrm{R}}$), heavy-quark mass value, and the uncertainty of the CTEQ6.6~\cite{Pumplin:2002vw} and CT14nlo~\cite{PhysRevD.93.033006} parton distribution functions, respectively.
Consistent with previous studies at lower collision energy~\cite{ALICE:2021mgk}, FONLL calculations describe the measured non-prompt D-meson production cross sections within uncertainties. In particular, the central values of the FONLL+PYTHIA~8 predictions are consistent with the data for the three D-meson species. The TAMU predictions using the statistical hadronisation approach for the abundances of different hadron species agree with the measured cross sections of non-prompt $\Dzero$ and $\Dplus$ mesons. However, they tend to overestimate the yield of non-prompt $\Ds$ mesons. The GM-VFNS calculations underestimate the measurements in the low-$\pt$ region, whereas at higher $\pt$ a better agreement is found.

%% file: results_DmesonRatios.tex
\subsection{Cross section ratios}
\label{sec:ratios}

\begin{figure}[tb!]
    \centering
    \includegraphics[width=0.49\textwidth]{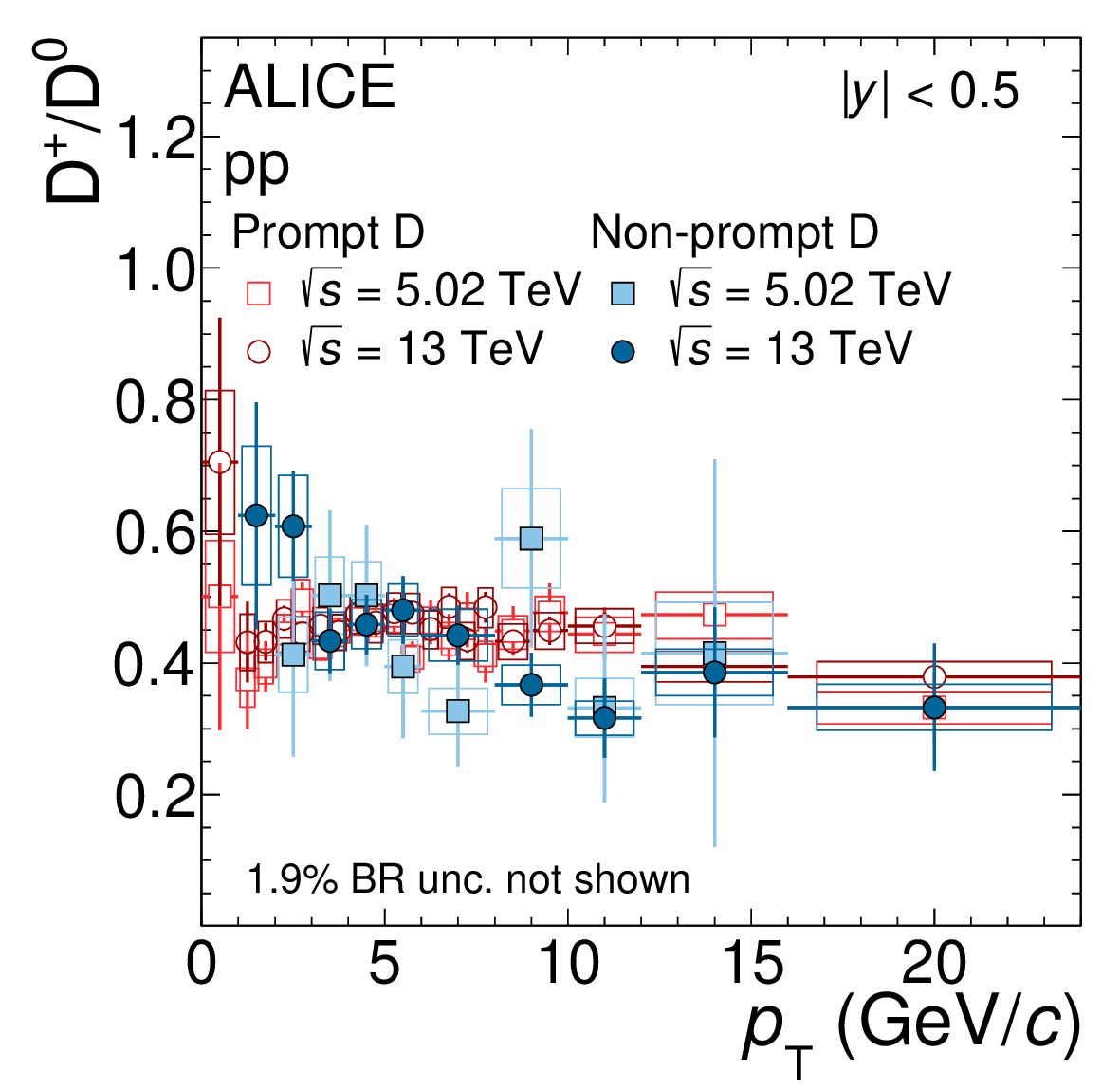}
    \includegraphics[width=0.49\textwidth]{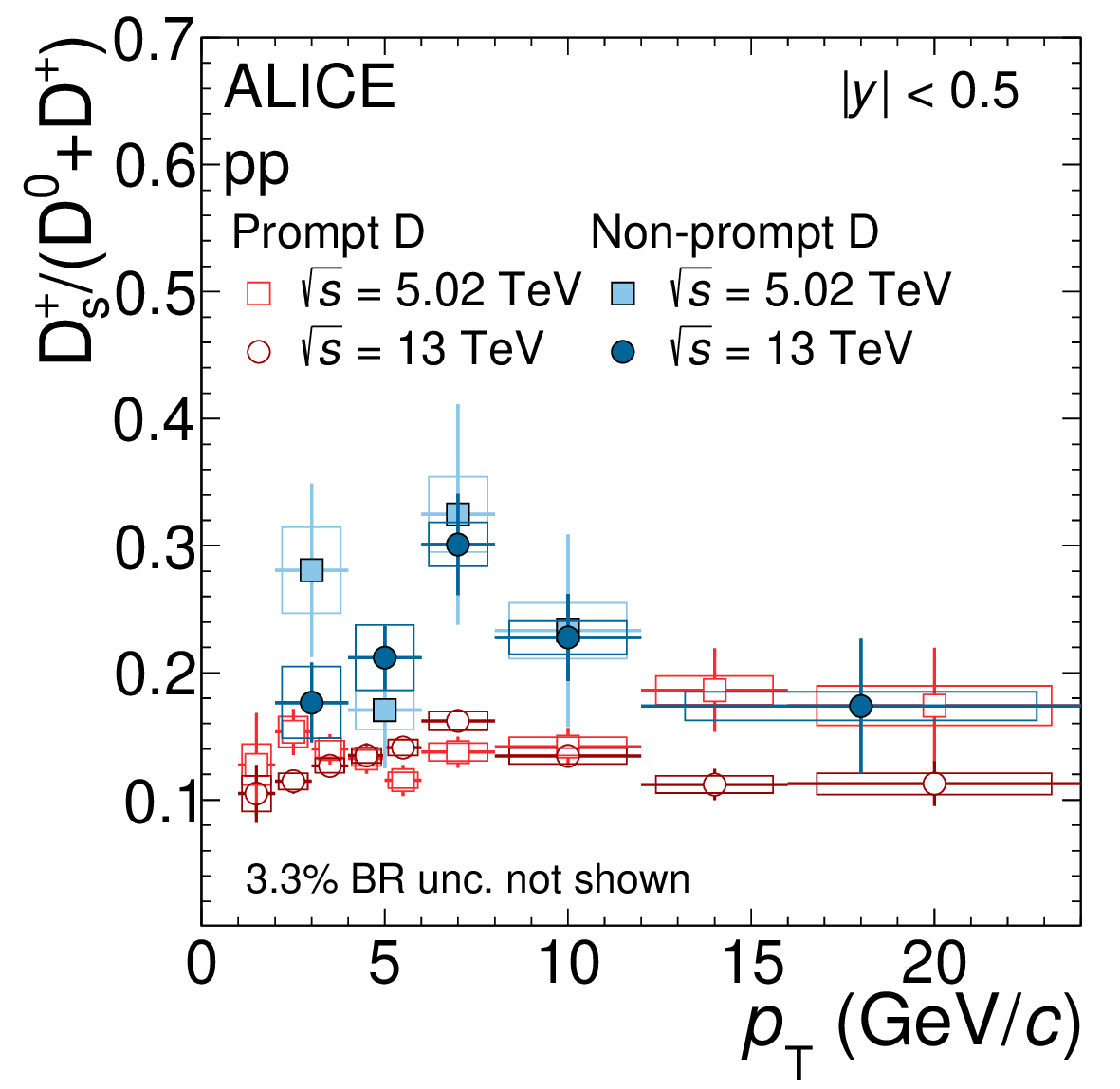}
    \includegraphics[width=0.49\textwidth]{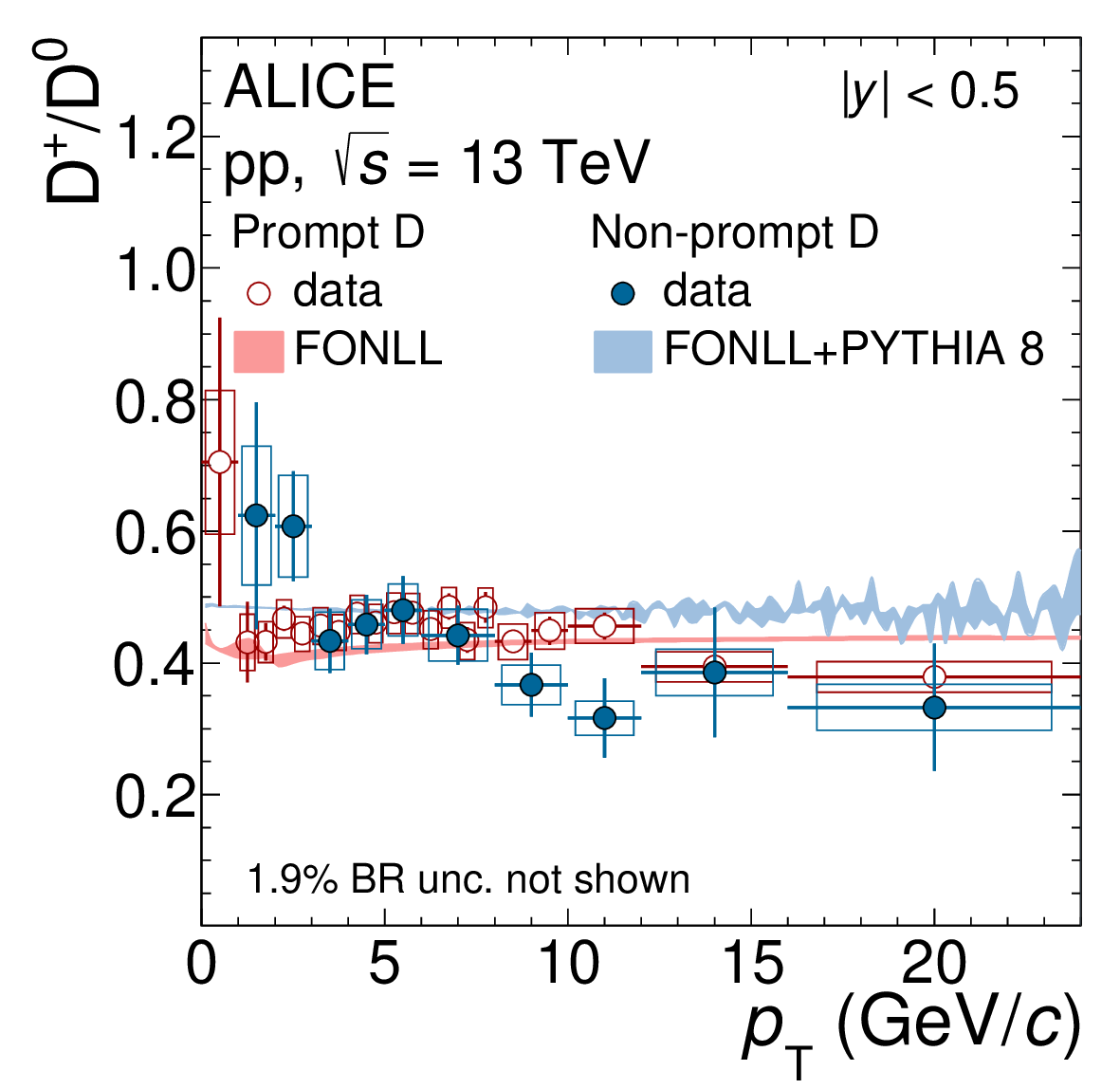}
    \includegraphics[width=0.49\textwidth]{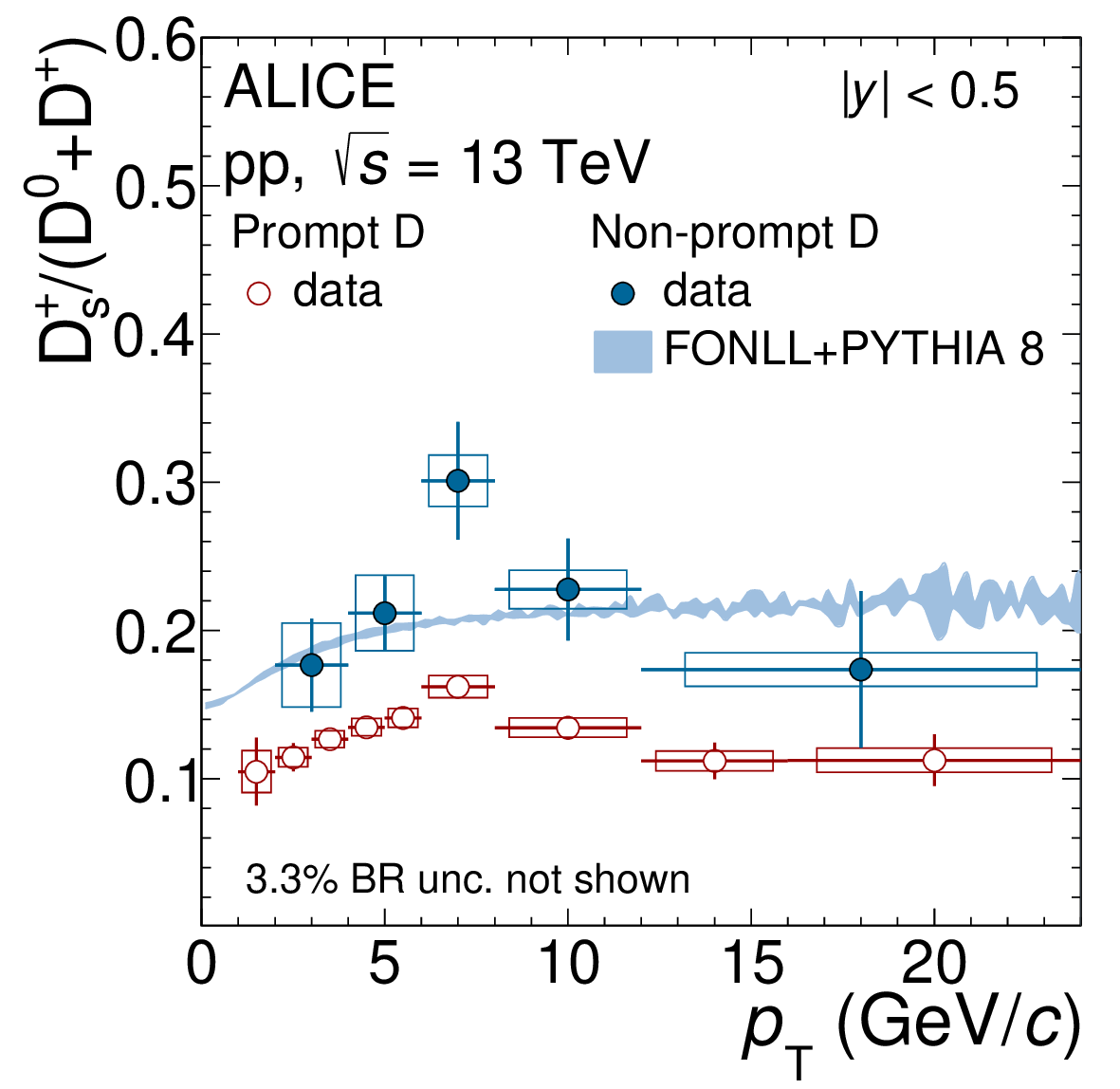}
    \caption{Top: ratios of prompt and non-prompt D-meson production cross sections as a function of $\pt$ in pp collisions at $\s = 5.02$~\cite{ALICE:2021mgk} and 13~\cite{ALICE:charm_pp13}~TeV. Bottom: ratios of prompt~\cite{ALICE:charm_pp13} and non-prompt D-meson production cross sections as a function of $\pt$ in pp collisions at $\s = 13$~TeV compared with FONLL+PYTHIA~8 predictions.
    The vertical bars and empty boxes report the statistical and systematic uncertainties, respectively.}
    \label{fig:NPDmesonRatios_5_13}
\end{figure}

The $\pt$-differential $\Dplus/\Dzero$ and $\Ds/(\Dzero+\Dplus)$ yield ratios are shown in the top row of Fig.~\ref{fig:NPDmesonRatios_5_13} for both non-prompt and prompt production in pp collisions at $\s = 5.02$ and $13$~TeV. All the systematic uncertainties are propagated in the ratios treating them as correlated among different D-meson species, except for the ones related to the raw-yield extraction, the selection efficiency, the raw non-prompt fraction estimation, and the BR, which are treated as uncorrelated. While the prompt $\Ds/(\Dzero+\Dplus)$ ratio~\cite{ALICE:charm_pp13} indicates a hint of an increase with increasing $\pt$ for $\pt$ below 8~GeV/$c$, no significant dependence is visible for the strange-to-non-strange ratio for non-prompt mesons and no firm conclusions can be drawn given the current experimental uncertainties. Also, no significant dependence on $\sqrt{s}$ is observed. This indicates that the ratios of fragmentation fractions of charm and beauty quarks into D mesons, as determined in pp collisions, exhibit no dependence on the collision energy, as detailed in Ref.~\cite{ALICE:charm_pp13}. 
The ratios are found to be compatible with those measured in $\ee$ collisions~\cite{HeavyFlavorAveragingGroup:2022wzx}, indicating that the fragmentation fractions of heavy quarks into mesons are independent of the collision system. Note that this does not apply to the baryon sector, where noticeable differences have been observed between $\ee$ and pp (or p--Pb) collisions~\cite{ALICE:2020wfu, ALICE:charm_pp13, ALICE:NPLc13TeV}.

In the bottom panels of Fig.~\ref{fig:NPDmesonRatios_5_13}, the ratios of prompt and non-prompt D-meson cross sections measured at $\s=13$~TeV are compared to the FONLL calculations for prompt D mesons and FONLL+PYTHIA~8 for the non-prompt ones. In the case of prompt $\Ds$ mesons, no FONLL prediction is currently available.
The theoretical predictions agree with the measured ratios in the $\pt$ range of the analyses.

From the measured non-prompt $\Ds / (\Dzero + \Dplus)$ ratios, it is possible to compute the fragmentation fraction ratio of beauty quarks into strange ($f_{\rm s}$) and non-strange ($f_{\rm u}$ and $f_{\rm d}$) B mesons at a centre-of-mass energy of \mbox{$\s =~13$~}TeV. 
It is important to consider that a significant portion of the non-prompt $\Ds$ mesons originates from decays of non-strange B mesons.
Therefore, a correction factor is applied to the $\pt$-differential non-prompt $\Ds / (\Dzero + \Dplus)$ ratio.
The correction factor is calculated from the FONLL+PYTHIA~8 predictions as
\begin{equation}
    \alpha_{\rm corr.}^{\rm FONLL + PYTHIA~8} = \left[ \frac{N (\Ds \leftarrow \Bs)}{N (\Ds \leftarrow \Hb)} \times \frac{N (\Dzero, \Dplus \leftarrow \Hb)}{N (\Dzero, \Dplus \leftarrow \rm B^{0, +})} \right]^{\rm FONLL + PYTHIA~8}\;,
    \label{eq:corr_fact_FF}
\end{equation}
where \mbox{$N (\Ds \leftarrow \Bs)$} is the number of $\Ds$ mesons produced in the decays of $\Bs$ mesons, \mbox{$N (\Dzero, \Dplus \leftarrow \rm B^{0, +})$} is the number of non-strange D mesons originating from non-strange B-meson decays, and \mbox{$N (\Ds \leftarrow \Hb)$} is the number of $\Ds$ from beauty-hadron decays. The correction factor is found to be around 0.5. Given that the majority of the
non-prompt $\Dzero$ and $\Dplus$ originate from non-strange B mesons, this indicates that about half of the non-prompt $\Ds$ mesons originate from decays of strange B mesons.

\begin{figure}[tb!]
    \centering
    \includegraphics[width=0.8\textwidth]{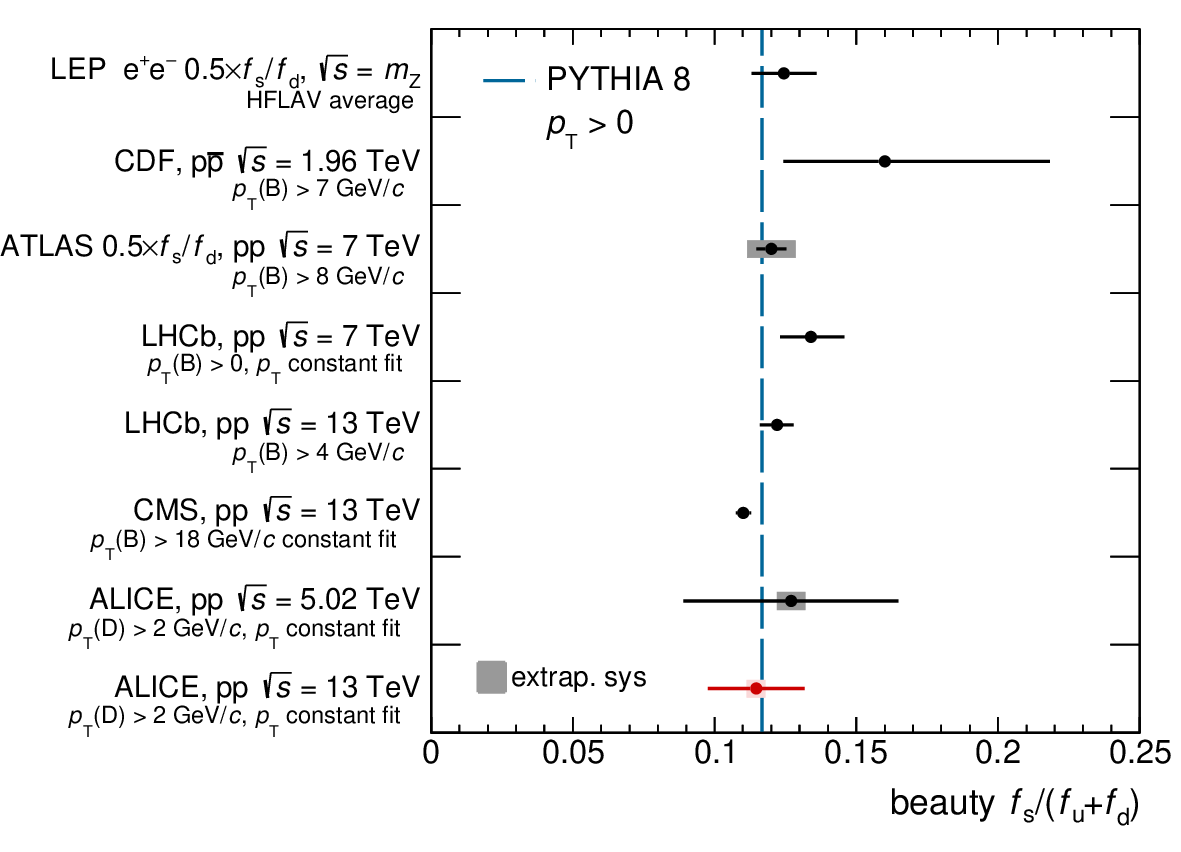}
    \caption{Beauty-quark fragmentation-fraction ratio $f_{\rm  s}/(f_{\rm u} + f_{\rm d})$ calculated from non-prompt D-meson measurements in pp collisions at $\s=13$~TeV compared
with previous measurements from CDF~\cite{CDF:2008yux}, ALICE~\cite{ALICE:2021mgk}, ATLAS~\cite{ATLAS:2015esn}, CMS~\cite{PhysRevLett.131.121901} and LHCb~\cite{LHCb:2011leg,LHCb:2019fns}, with the average of LEP measurements~\cite{HeavyFlavorAveragingGroup:2022wzx}, and with the results of PYTHIA 8.243 simulations with Monash-13 tune~\cite{Skands:2014pea}.}
    \label{fig:FFs}
\end{figure}

The ratio of the beauty-quark fragmentation fractions $\fsratio$ is computed as
\begin{equation}
    \left( \frac{f_{\rm  s}}{f_{\rm u} + f_{\rm d} } \right)_{\rm beauty} = \alpha_{\rm corr.}^{\rm FONLL + PYTHIA~8} \times \left( \frac{\Ds}{\Dzero + \Dplus} \right)_{\rm non\mbox{-}prompt},
    \label{eq:FF}
\end{equation}
where, in addition to the uncertainty related to the $\Ds/(\Dzero + \Dplus)$ measurement, the uncertainty on the correction factor $\alpha_{\rm corr.}^{\rm FONLL + PYTHIA~8}$ was considered. The latter was estimated by varying the beauty-quark fragmentation fractions and the branching ratios adopted for the ${\rm h_b \rightarrow D+X}$ decays in PYTHIA~8.243 besides the uncertainties of FONLL, as discussed in detail in Ref.~\cite{ALICE:2021mgk}.
The ratio $\fsratio$ was calculated in the $\pt$ intervals of the non-prompt $\Ds / (\Dzero + \Dplus)$ ratio measurement and it was found to be independent of $\pt$ within the experimental uncertainties.
It was fitted with a constant function to obtain the $\pt$-integrated fragmentation fraction. The result is:
\begin{equation}
    \left( \frac{f_{\rm  s}}{f_{\rm u} + f_{\rm d} } \right)_{\rm beauty} = 0.114 \pm 0.016~{\rm (stat.)} \pm 0.006~{\rm (syst.)} \pm 0.003~{\rm (BR)} \pm 0.003~{\rm (extrap.)},
    \label{eq:FFresult}
\end{equation}
where ``stat.'' denotes the statistical uncertainty, ``BR'' is the uncertainty on the BR of the considered decays, and ``extrap.'' denotes the uncertainty due to the $\pt$ extrapolation. The systematic uncertainty is denoted as ``syst.'' and it takes into account the yield extraction, track reconstruction and selection, D meson selection, non-prompt fraction, and simulated $\pt$ shape. 
In Fig.~\ref{fig:FFs}, the fragmentation fraction ratio is compared to previous measurements in pp and ${\rm p\overline{p}}$ collisions at different $\s$ values from the CDF~\cite{CDF:2008yux}, ALICE~\cite{ALICE:2021mgk}, ATLAS~\cite{ATLAS:2015esn}, and LHCb~\cite{LHCb:2011leg,LHCb:2019fns} Collaborations. 
The ATLAS measurement of $f_{\rm s}/f_{\rm d}$ was downscaled by a factor of two, assuming isospin symmetry for the u and d quarks (i.e.\ $f_{\rm u} = f_{\rm d}$). Figure~\ref{fig:FFs} also shows that the beauty-quark fragmentation fractions extracted from the measurements in pp (${\rm p\overline{p}}$) collisions are compatible with those computed by the HFLAV Collaboration~\cite{HeavyFlavorAveragingGroup:2022wzx} using measurements in e$^+$e$^-$ collisions at LEP. Similarly to the ATLAS measurement, the HFLAV results were scaled by a factor of two.
The larger data sample collected by ALICE at $\sqrt{s} = 13$~TeV, compared to lower collision energies, enables a significant improvement in the precision of the measured beauty-quark fragmentation fraction ratio. Notably, the fragmentation fraction ratio for beauty quarks presented in this work is comparable to the one measured for charm quarks, which was determined to be $0.116 \pm 0.011~{\rm (stat.)} \pm 0.009~{\rm (syst.)} \pm 0.003~{\rm (BR)}$~\cite{ALICE:charm_pp13} at a centre-of-mass energy of 13~TeV and $0.136 \pm 0.005~{\rm (stat.)}  \pm 0.006 ~\rm (syst.) \pm 0.005 ~{\rm (BR)}$~\cite{ALICE:2021mgk} at 5.02~TeV. It is also consistent with the value of the strange to non-strange ratio for light-flavour particles predicted by the statistical hadronisation model~\cite{Braun-Munzinger:2001alt}, which is about 0.1, and with the outcome of PYTHIA~8.243 simulations with the Monash-13 tune~\cite{Skands:2014pea}.

\begin{figure}[tb!]
    \centering
    \includegraphics[width=1.0\textwidth]{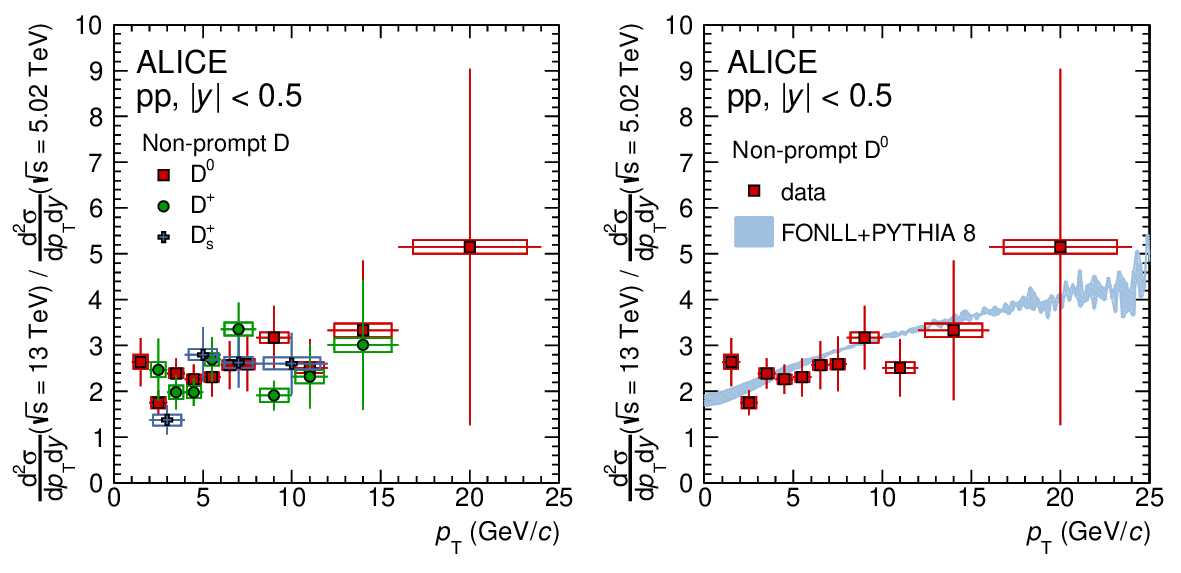}
    \caption{Ratios of non-prompt D-meson production cross sections at $\s = 13$ and $5.02$~TeV~\cite{ALICE:2021mgk} as a function of $\pt$. Left: comparison of measurements for $\Dzero$, $\Dplus$, and $\Ds$ mesons. Right: comparison of $\Dzero$ results with FONLL+PYTHIA~8 calculations. The vertical bars and empty boxes report the statistical and systematic uncertainties, respectively.}
    \label{fig:MesonToMesonRatio-13to5}
\end{figure}

To further explore the relationship of non-prompt D-meson production with the \pp collision centre-of-mass energy, the ratios between the $\pt$-differential non-prompt D-meson cross sections at $\s = 13$ and $5.02$~TeV~\cite{ALICE:2019nxm, ALICE:2021mgk} were computed for $\Dzero$, $\Dplus$, and $\Ds$ mesons. The results are shown in the left panel of Fig.~\ref{fig:MesonToMesonRatio-13to5}. The systematic uncertainties were propagated treating them as uncorrelated between the two collision energies, with the exception of those related to the non-prompt fraction estimation and the branching ratio, which were treated as fully correlated. The ratios for the different D-meson species are compatible within the experimental uncertainties and share a common trend in $\pt$. The ratios hint at a common increase with increasing $\pt$, as also reported for charm-hadron ratios~\cite{ALICE:charm_pp13} and the equivalent ratio for $\Bplus$ mesons in the rapidity interval $2.0 < y < 4.5$~\cite{LHCb:2017vec}. These findings suggest that there is a similar hardening of the $\pt$-differential production cross section of heavy-flavour hadrons with increasing $\s$, which is independent of the species or origin of the hadron. In addition, a comparison was made between the measured 13-to-5.02 TeV ratio for non-prompt $\Dzero$ and the FONLL+PYTHIA~8 predictions, which is shown in the right panel of Fig.~\ref{fig:MesonToMesonRatio-13to5}. The non-prompt $\Dzero$ results were chosen for this comparison due to the larger $\pt$ coverage of the measurement and the higher precision compared to the other D meson species. The theoretical calculations also indicate an increasing trend of the non-prompt $\Dzero$ ratio as a function of $\pt$, and they reproduce the measurement within uncertainties. However, the current experimental uncertainties do not allow us to draw a firm conclusion on the dependence of the ratios on the D-meson $\pt$.

%% file: results_FFbbbar.tex
\subsection{Beauty-quark production in pp collisions at $\mathbf{\sqrt{\textit{s}}=13}$ TeV}
\label{sec:bbar}

The production cross section of $\bbbar$ pairs per unit of rapidity at midrapidity was computed independently for the non-prompt $\Dzero$, $\Dplus$, $\Ds$ mesons, and $\Lambdac$ baryons taken from Ref.~\cite{ALICE:NPLc13TeV} by combining two ingredients: (i) the measured visible cross sections for that specific species and (ii) the extrapolation factor to the $\bbbar$-pair cross section ($\alpha_{\rm extrap}^{\bbbar}$) taken from FONLL calculations. These two quantities were estimated following the procedure presented in Refs.~\cite{ALICE:2021mgk, ALICE:NPLc13TeV} and summarised below. Finally, the total $\bbbar$ cross section per unit of rapidity at midrapidity was estimated as the weighted average of the single $\bbbar$ cross sections of the various charm hadron species, $\rm h_{c}$, adopting as weights the inverse of the quadratic sum of the absolute statistical and uncorrelated systematic uncertainties, encompassing the systematic uncertainty associated with the raw-yield extraction and the non-prompt fraction estimation.

The visible cross sections of the different non-prompt charm-hadron species were obtained by integrating the $\pt$-differential cross section in the measured $\pt$ interval. In this calculation, the statistical uncertainty and the systematic uncertainty on the raw-yield extraction were treated as uncorrelated among the different $\pt$ intervals. All the other sources were considered fully correlated. 
The extrapolation factor was computed for each measured non-prompt charm-hadron state from FONLL+PYTHIA~8 predictions, as the ratio of the beauty cross section and the visible cross sections of the measured non-prompt charm-hadron states in the $\pt$ range of the analyses, as:

\begin{equation}
    \centering
    \alpha_{\mathrm {extrap.}}^{\mathrm{ b\overline{b}}} = \frac{\mathrm{d} \sigma^{\mathrm{FONLL}}_{\bbbar}/\mathrm{d} y|_{|y| < 0.5}}{ \mathrm{d} \sigma^{\mathrm{FONLL+PYTHIA~8}}_{\mathrm{h_{c}}\leftarrow\mathrm{b}}/\mathrm{d} y|_{|y|<0.5} (\pt^{\mathrm{min~h_{c}}} < \pt < \pt^{\mathrm{max~h_{c}}})}\;.
    \label{eq:alpha_bbbar}
\end{equation}

The systematic uncertainty on the extrapolation factor was determined by considering different sources of uncertainty including the ones associated with FONLL calculations, the uncertainties on the fragmentation fractions, and the uncertainties on the BR as was done for the estimation of the beauty-quark fragmentation fractions described in Section~\ref{sec:ratios}. Table~\ref{tab:crossSecD} reports the kinematic range of the analyses together with the measured visible cross sections, the extrapolation factor, 
and the predictions of FONLL+PYTHIA~8 calculations for each non-prompt charm-hadron species~\cite{ALICE:NPLc13TeV}.

Two additional corrections were considered, following the procedure of Refs~\cite{ALICE:charm_pp13, ALICE:2021mgk, ALICE:NPLc13TeV}. The first correction factor accounts for the different rapidity distributions of beauty quarks and beauty hadrons, while the second correction accounts for the different rapidity distributions of $\bbbar$ pairs and beauty quarks. The first correction factor was estimated to be unity within the relevant rapidity range, based on FONLL calculations. An uncertainty of 1\% was derived from the deviation between FONLL and PYTHIA~8.243. The second correction factor was computed from the rapidity distributions of b quarks and $\bbbar$ pairs obtained with POWHEG~\cite{Alioli:2010xd} simulations. The resulting value of the second correction factor is $1.06 \pm 0.01$ within the range $|y|<0.5$, where the uncertainty was derived by varying the factorisation and renormalisation scales in the POWHEG calculation.

\begin{table}[tb!]
    \renewcommand*{\arraystretch}{1.4}
    \centering
    \caption{Measured visible cross section (${\rm d}\sigma^{\rm vis.}/{\rm d}y|_{|y|<0.5}$), extrapolation factor ($\alpha_{\mathrm {extrap.}}^{\mathrm{ b\overline{b}}}$), and FONLL+PYTHIA~8 predictions of non-prompt charm-hadrons in \pp collisions at $\s=13$ TeV at midrapidity.}
    \begin{tabular}{lcccc}
       \toprule
         $\rm h_{c}$ & Kinematic range& \multicolumn{1}{c}{${\rm d}\sigma^{\rm vis.}/{\rm d}y|_{|y|<0.5}$} & \multicolumn{1}{c}{$\alpha_{\mathrm {extrap.}}^{\mathrm{ b\overline{b}}}$}
         & FONLL+PYTHIA~8\\
          & (GeV/$c$) & ($\upmu$b) &  & ($\upmu$b) \\
        \midrule

        $\Dzero$
        & 1 $< \pt <$ 24 
        & 33.3 $\pm$ 1.2 (stat.) $\pm$ 2.5 (syst.) 
        & 1.241 $^{+0.009}_{-0.047}$ 
        & 30.7 $^{+13.7}_{-12.2}$
        \\

        $\Dplus$
        & 1 $< \pt <$ 24
        & 18.4 $\pm$ 2.4 (stat.) $\pm$ 2.3 (syst.)
        & 1.243 $^{+0.009}_{-0.048}$
        & 14.8 $^{+6.6}_{-5.9}$
        \\

        $\Ds$ 
        & 2 $< \pt <$ 24
        & 5.8 $\pm$ 0.6 (stat.) $\pm$ 0.4 (syst.)
        & 1.858 $^{+0.037}_{-0.170}$
        & 6.3 $^{+2.7}_{-2.3}$
        \\

        $\Lambdac$~\cite{ALICE:NPLc13TeV} 
        & 2 $< \pt <$ 24
        & 10.5 $\pm$ 1.3 (stat.) $\pm$ 0.9 (syst.)
        & 1.847 $^{+0.108}_{-0.152}$
        & 5.9 $^{+3.3}_{-2.4}$
        \\
        \bottomrule
    \end{tabular}
    \label{tab:crossSecD}
\end{table}

\begin{figure}[tb!]
    \centering
    \includegraphics[width=0.45\textwidth]{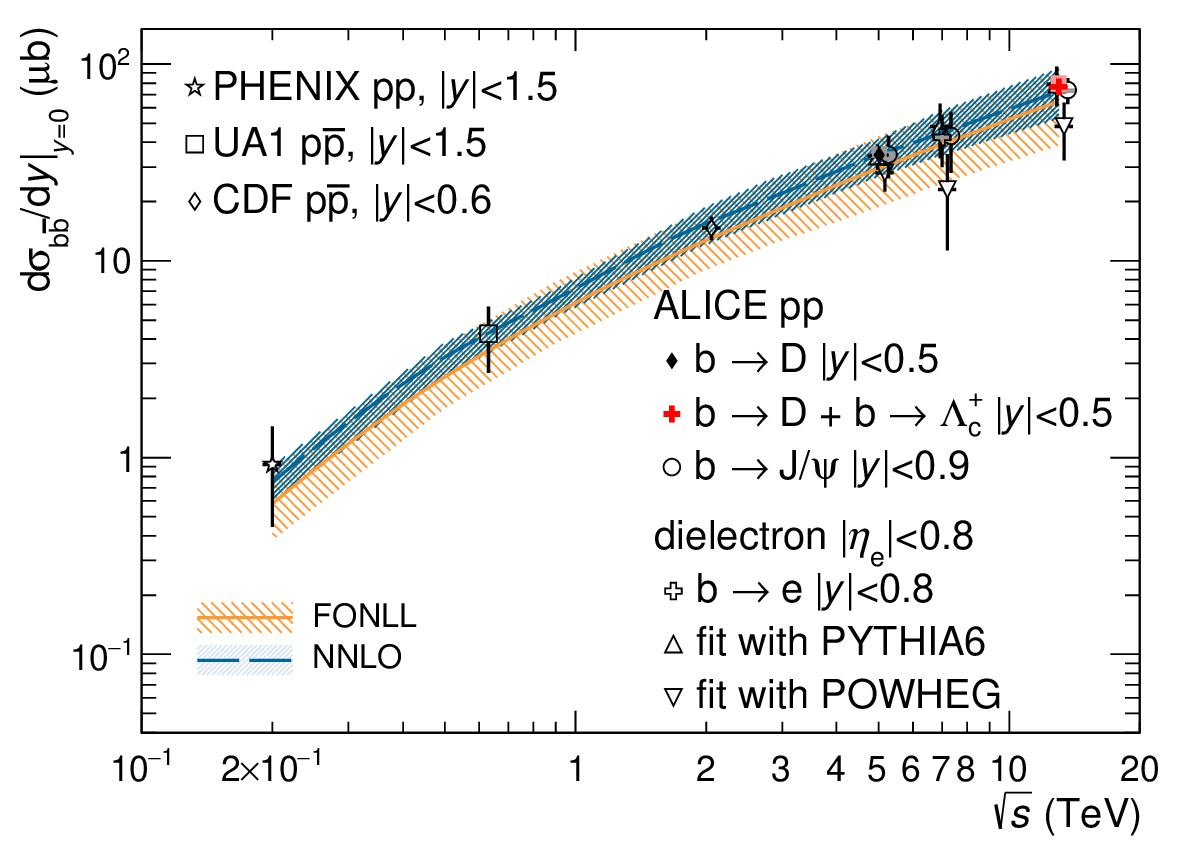}
    \includegraphics[width=0.49\textwidth]{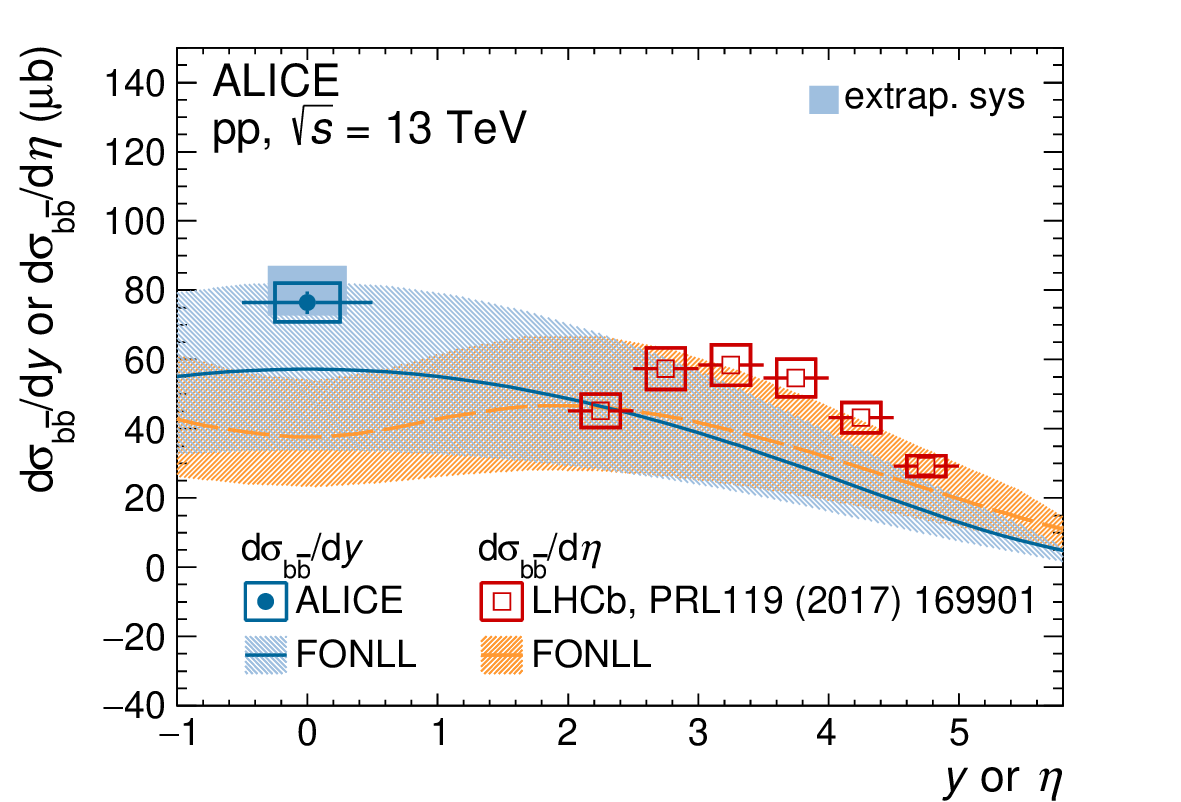}
    \caption{Left: beauty-quark production cross section per unit of rapidity at midrapidity as a function of the centre-of-mass energy measured in pp and ${\rm p\overline{p}}$ collisions by the ALICE~\cite{ALICE:2021edd,ALICE:2012vpz,ALICE:2012acz,ALICE:2018fvj,ALICE:2018gev,ALICE:2020mfy} and PHENIX~\cite{PHENIX:2009dpd} Collaborations, and the CDF~\cite{CDF:2004jtw} and UA1~\cite{UA1:1990vvp} Collaborations. The solid and dashed lines, accompanied by shaded bands, represent the central values and the associated uncertainties predicted by FONLL~\cite{Cacciari:1998it, Cacciari:2001td, Cacciari:2012ny} and NNLO~\cite{Catani:2020kkl} calculations, respectively. Right: beauty-quark production cross section per unit of (pseudo)rapidity as a function of (pseudo)rapidity measured by the ALICE Collaboration (LHCb Collaboration~\cite{LHCb:2016qpe}). The solid and dashed lines, accompanied by a shaded band, represent the central values and the associated uncertainties predicted by FONLL~\cite{Cacciari:1998it, Cacciari:2001td, Cacciari:2012ny} as a function of $y$ and $\eta$, respectively.
     The vertical bars and boxes report the statistical and systematic uncertainties, respectively.}
    \label{fig:bb_vs_srts}
\end{figure}

The measured $\bbbar$ production cross section per unit of rapidity at midrapidity in pp collisions at $\s = 13$~TeV is
\begin{equation}{
\left.\frac{{\rm d}\sigma_{\bbbar}}{{\rm d}y}\right|_{|y|<0.5}^{\text{\pp}, \s=13\text{~TeV}} = 75.2 \pm 3.2 (\mathrm{stat.}) \pm 5.2 (\mathrm{syst.})^{+12.3}_{-3.2} (\mathrm{extrap.})\text{ } \rm \upmu b \;,
    \label{eq:ccbarpp13TeV}
}
\end{equation}
where the uncertainty on the extrapolation of the cross section is reported separately and denoted as ``extrap.''.  
The left panel of Fig.~\ref{fig:bb_vs_srts} reports a compilation of $\bbbar$ cross section measurements in pp collisions from the ALICE~\cite{ALICE:2021edd,ALICE:2012vpz,ALICE:2012acz,ALICE:2018fvj,ALICE:2018gev,ALICE:2020mfy} and PHENIX~\cite{PHENIX:2009dpd} Collaborations and in ${\rm p\overline{p}}$ from the CDF~\cite{CDF:2004jtw} and UA1~\cite{UA1:1990vvp} Collaborations, as a function of $\s$. The experimental results are compared with the predictions from FONLL and NNLO calculations. The measured $\bbbar$ cross sections and their dependence with $\s$ are described by the pQCD calculations within the theoretical uncertainties, especially in the case of NNLO calculations that show smaller uncertainties compared to FONLL calculations. The central values of the NNLO calculations are closer to the data, as expected by the higher perturbative accuracy.

Finally, the dependence of the $\bbbar$ production cross section on the rapidity interval was investigated. The right panel of Fig.~\ref{fig:bb_vs_srts} reports the beauty cross section at midrapidity presented in this work, along with the measurements performed in pseudorapidity intervals by the LHCb Collaboration at forward rapidity~\cite{LHCb:2016qpe}. The experimental results are compared to FONLL predictions, which are shown both for ${\rm d}\sigma_\bbbar/{\rm d}y$ and ${\rm d}\sigma_\bbbar/{\rm d}\eta$, to match the observables reported by ALICE and LHCb, respectively. The measured $\bbbar$ cross sections generally lie close to the upper boundary of the FONLL theoretical uncertainty band, except for the LHCb data point in $2<\eta<2.5$ which is compatible with the central value of the predictions.

%% file: summary.tex
\section{Summary}
\label{sec:summary}
The $\pt$-differential production cross sections of non-prompt D mesons were measured at midrapidity in pp collisions at $\s = 13$~TeV. The measurement of the non-prompt $\Dzero$ meson was performed in the $1 < \pt < 24$~GeV/$c$ range with a finer granularity in $\pt$ compared to previously published measurements~\cite{ALICE:NPLc13TeV}. The measurements of the non-prompt $\Dplus$ and $\Ds$ mesons were performed for the first time at this energy and cover the $1 < \pt < 24$~GeV/$c$ and $2 < \pt < 24$~GeV/$c$ ranges, respectively. The results were compared with FONLL and GM-VFNS pQCD calculations as well as with predictions from the TAMU model. 
A good agreement was found considering the current experimental uncertainties.

The non-prompt $\Dplus$/$\Dzero$ and $\Ds$/($\Dzero$+$\Dplus$) $\pt$-differential production ratios were studied at $\s = 13$~TeV and were compared to the 5.02 TeV results. No significant dependence on the collision energy nor on the D-meson $\pt$ was observed. A comparison was made with the predictions based on FONLL+PYTHIA~8, which describe the measured ratios well. These results were further employed to test the universality of the fragmentation function of beauty quarks in pp collisions by measuring the fraction of beauty quarks fragmenting into strange mesons relative to those fragmenting into non-strange mesons. The result is compatible with previously published measurements in $\ee$, $\ppbar$, and pp collisions at different energies and with the analogous measurement in the charm sector~\cite{ALICE:charm_pp13}. Our results are consistent with the scenario in which the ratio of the fragmentation fraction of charm and beauty quarks into D mesons is universal, i.e.\ independent of the collision system and energy, even though violations of this universality are observed when considering heavy-flavour baryons~\cite{LHCb:2011leg,ALICE:2020wfu,ALICE:2020wla,ALICE:2022ych,Acharya:2021vpo,Acharya:2021vjp,LHCb:2019fns}. In addition, the ratio of the non-prompt D meson production between $\s=13$ and 5.02~TeV was measured for the three meson species. A hint of a common increase with increasing $\pt$ was observed for all species, similar to what was already reported in analogous measurements of charm hadrons~\cite{ALICE:charm_pp13}, even though the current experimental uncertainty does not allow firm conclusions to be drawn. Predictions based on FONLL+PYTHIA~8 calculations qualitatively describe these ratios.

Finally, the total $\bbbar$ production cross section at midrapidity per unity of rapidity in pp collisions at $\s = 13$~TeV was determined. This result supersedes the previous measurement presented in Ref.~\cite{ALICE:NPLc13TeV} in terms of precision. The measured $\bbbar$ production cross section lies on the upper edge of the theoretical uncertainties of the predictions from FONLL and NNLO pQCD calculations.

%% file: fa_2024-02-15_Opt_C.tex

The ALICE Collaboration would like to thank all its engineers and technicians for their invaluable contributions to the construction of the experiment and the CERN accelerator teams for the outstanding performance of the LHC complex.
The ALICE Collaboration gratefully acknowledges the resources and support provided by all Grid centres and the Worldwide LHC Computing Grid (WLCG) collaboration.
The ALICE Collaboration acknowledges the following funding agencies for their support in building and running the ALICE detector:
A. I. Alikhanyan National Science Laboratory (Yerevan Physics Institute) Foundation (ANSL), State Committee of Science and World Federation of Scientists (WFS), Armenia;
Austrian Academy of Sciences, Austrian Science Fund (FWF): [M 2467-N36] and Nationalstiftung f\"{u}r Forschung, Technologie und Entwicklung, Austria;
Ministry of Communications and High Technologies, National Nuclear Research Center, Azerbaijan;
Conselho Nacional de Desenvolvimento Cient\'{\i}fico e Tecnol\'{o}gico (CNPq), Financiadora de Estudos e Projetos (Finep), Funda\c{c}\~{a}o de Amparo \`{a} Pesquisa do Estado de S\~{a}o Paulo (FAPESP) and Universidade Federal do Rio Grande do Sul (UFRGS), Brazil;
Bulgarian Ministry of Education and Science, within the National Roadmap for Research Infrastructures 2020-2027 (object CERN), Bulgaria;
Ministry of Education of China (MOEC) , Ministry of Science \& Technology of China (MSTC) and National Natural Science Foundation of China (NSFC), China;
Ministry of Science and Education and Croatian Science Foundation, Croatia;
Centro de Aplicaciones Tecnol\'{o}gicas y Desarrollo Nuclear (CEADEN), Cubaenerg\'{\i}a, Cuba;
Ministry of Education, Youth and Sports of the Czech Republic, Czech Republic;
The Danish Council for Independent Research | Natural Sciences, the VILLUM FONDEN and Danish National Research Foundation (DNRF), Denmark;
Helsinki Institute of Physics (HIP), Finland;
Commissariat \`{a} l'Energie Atomique (CEA) and Institut National de Physique Nucl\'{e}aire et de Physique des Particules (IN2P3) and Centre National de la Recherche Scientifique (CNRS), France;
Bundesministerium f\"{u}r Bildung und Forschung (BMBF) and GSI Helmholtzzentrum f\"{u}r Schwerionenforschung GmbH, Germany;
General Secretariat for Research and Technology, Ministry of Education, Research and Religions, Greece;
National Research, Development and Innovation Office, Hungary;
Department of Atomic Energy Government of India (DAE), Department of Science and Technology, Government of India (DST), University Grants Commission, Government of India (UGC) and Council of Scientific and Industrial Research (CSIR), India;
National Research and Innovation Agency - BRIN, Indonesia;
Istituto Nazionale di Fisica Nucleare (INFN), Italy;
Japanese Ministry of Education, Culture, Sports, Science and Technology (MEXT) and Japan Society for the Promotion of Science (JSPS) KAKENHI, Japan;
Consejo Nacional de Ciencia (CONACYT) y Tecnolog\'{i}a, through Fondo de Cooperaci\'{o}n Internacional en Ciencia y Tecnolog\'{i}a (FONCICYT) and Direcci\'{o}n General de Asuntos del Personal Academico (DGAPA), Mexico;
Nederlandse Organisatie voor Wetenschappelijk Onderzoek (NWO), Netherlands;
The Research Council of Norway, Norway;
Commission on Science and Technology for Sustainable Development in the South (COMSATS), Pakistan;
Pontificia Universidad Cat\'{o}lica del Per\'{u}, Peru;
Ministry of Education and Science, National Science Centre and WUT ID-UB, Poland;
Korea Institute of Science and Technology Information and National Research Foundation of Korea (NRF), Republic of Korea;
Ministry of Education and Scientific Research, Institute of Atomic Physics, Ministry of Research and Innovation and Institute of Atomic Physics and Universitatea Nationala de Stiinta si Tehnologie Politehnica Bucuresti, Romania;
Ministry of Education, Science, Research and Sport of the Slovak Republic, Slovakia;
National Research Foundation of South Africa, South Africa;
Swedish Research Council (VR) and Knut \& Alice Wallenberg Foundation (KAW), Sweden;
European Organization for Nuclear Research, Switzerland;
Suranaree University of Technology (SUT), National Science and Technology Development Agency (NSTDA) and National Science, Research and Innovation Fund (NSRF via PMU-B B05F650021), Thailand;
Turkish Energy, Nuclear and Mineral Research Agency (TENMAK), Turkey;
National Academy of  Sciences of Ukraine, Ukraine;
Science and Technology Facilities Council (STFC), United Kingdom;
National Science Foundation of the United States of America (NSF) and United States Department of Energy, Office of Nuclear Physics (DOE NP), United States of America.
In addition, individual groups or members have received support from:
Czech Science Foundation (grant no. 23-07499S), Czech Republic;
European Research Council (grant no. 950692), European Union;
ICSC - Centro Nazionale di Ricerca in High Performance Computing, Big Data and Quantum Computing, European Union - NextGenerationEU;
Academy of Finland (Center of Excellence in Quark Matter) (grant nos. 346327, 346328), Finland.

%% file: 2024-02-15-Alice_Authorlist_2024-02-15_Opt_C.tex
\begin{flushleft} 
\small

S.~Acharya\,\orcidlink{0000-0002-9213-5329}\,$^{\rm 128}$, 
D.~Adamov\'{a}\,\orcidlink{0000-0002-0504-7428}\,$^{\rm 87}$, 
A.~Agarwal$^{\rm 136}$, 
G.~Aglieri Rinella\,\orcidlink{0000-0002-9611-3696}\,$^{\rm 33}$, 
L.~Aglietta$^{\rm 25}$, 
M.~Agnello\,\orcidlink{0000-0002-0760-5075}\,$^{\rm 30}$, 
N.~Agrawal\,\orcidlink{0000-0003-0348-9836}\,$^{\rm 26}$, 
Z.~Ahammed\,\orcidlink{0000-0001-5241-7412}\,$^{\rm 136}$, 
S.~Ahmad\,\orcidlink{0000-0003-0497-5705}\,$^{\rm 16}$, 
S.U.~Ahn\,\orcidlink{0000-0001-8847-489X}\,$^{\rm 72}$
% ...

\section*{Affiliation Notes}

$^{\rm I}$ Deceased\\
$^{\rm II}$ Also at: Max-Planck-Institut fur Physik, Munich, Germany\\
$^{\rm III}$ Also at: Italian National Agency for New Technologies, Energy and Sustainable Economic Development (ENEA), Bologna, Italy\\
$^{\rm IV}$ Also at: Dipartimento DET del Politecnico di Torino, Turin, Italy\\
$^{\rm V}$ Also at: Yildiz Technical University, Istanbul, T\"{u}rkiye\\
$^{\rm VI}$ Also at: Department of Applied Physics, Aligarh Muslim University, Aligarh, India\\
$^{\rm VII}$ Also at: Institute of Theoretical Physics, University of Wroclaw, Poland\\
$^{\rm VIII}$ Also at: An institution covered by a cooperation agreement with CERN\\

\section*{Collaboration Institutes}

$^{1}$ A.I. Alikhanyan National Science Laboratory (Yerevan Physics Institute) Foundation, Yerevan, Armenia\\
$^{2}$ AGH University of Krakow, Cracow, Poland\\
$^{3}$ Bogolyubov Institute for Theoretical Physics, National Academy of Sciences of Ukraine, Kiev, Ukraine\\
$^{4}$ Bose Institute, Department of Physics  and Centre for Astroparticle Physics and Space Science (CAPSS), Kolkata, India\\
$^{5}$ California Polytechnic State University, San Luis Obispo, California, United States\\
$^{6}$ Central China Normal University, Wuhan, China\\
$^{7}$ Centro de Aplicaciones Tecnol\'{o}gicas y Desarrollo Nuclear (CEADEN), Havana, Cuba\\
$^{8}$ Centro de Investigaci\'{o}n y de Estudios Avanzados (CINVESTAV), Mexico City and M\'{e}rida, Mexico\\
$^{9}$ Chicago State University, Chicago, Illinois, United States\\
$^{10}$ China Institute of Atomic Energy, Beijing, China\\
$^{11}$ China University of Geosciences, Wuhan, China\\
$^{12}$ Chungbuk National University, Cheongju, Republic of Korea\\
$^{13}$ Comenius University Bratislava, Faculty of Mathematics, Physics and Informatics, Bratislava, Slovak Republic\\
$^{14}$ COMSATS University Islamabad, Islamabad, Pakistan\\
$^{15}$ Creighton University, Omaha, Nebraska, United States\\
$^{16}$ Department of Physics, Aligarh Muslim University, Aligarh, India\\
$^{17}$ Department of Physics, Pusan National University, Pusan, Republic of Korea\\
$^{18}$ Department of Physics, Sejong University, Seoul, Republic of Korea\\
$^{19}$ Department of Physics, University of California, Berkeley, California, United States\\
$^{20}$ Department of Physics, University of Oslo, Oslo, Norway\\
$^{21}$ Department of Physics and Technology, University of Bergen, Bergen, Norway\\
$^{22}$ Dipartimento di Fisica, Universit\`{a} di Pavia, Pavia, Italy\\
$^{23}$ Dipartimento di Fisica dell'Universit\`{a} and Sezione INFN, Cagliari, Italy\\
$^{24}$ Dipartimento di Fisica dell'Universit\`{a} and Sezione INFN, Trieste, Italy\\
$^{25}$ Dipartimento di Fisica dell'Universit\`{a} and Sezione INFN, Turin, Italy\\
$^{26}$ Dipartimento di Fisica e Astronomia dell'Universit\`{a} and Sezione INFN, Bologna, Italy\\
$^{27}$ Dipartimento di Fisica e Astronomia dell'Universit\`{a} and Sezione INFN, Catania, Italy\\
$^{28}$ Dipartimento di Fisica e Astronomia dell'Universit\`{a} and Sezione INFN, Padova, Italy\\
$^{29}$ Dipartimento di Fisica `E.R.~Caianiello' dell'Universit\`{a} and Gruppo Collegato INFN, Salerno, Italy\\
$^{30}$ Dipartimento DISAT del Politecnico and Sezione INFN, Turin, Italy\\
$^{31}$ Dipartimento di Scienze MIFT, Universit\`{a} di Messina, Messina, Italy\\
$^{32}$ Dipartimento Interateneo di Fisica `M.~Merlin' and Sezione INFN, Bari, Italy\\
$^{33}$ European Organization for Nuclear Research (CERN), Geneva, Switzerland\\
$^{34}$ Faculty of Electrical Engineering, Mechanical Engineering and Naval Architecture, University of Split, Split, Croatia\\
$^{35}$ Faculty of Engineering and Science, Western Norway University of Applied Sciences, Bergen, Norway\\
$^{36}$ Faculty of Nuclear Sciences and Physical Engineering, Czech Technical University in Prague, Prague, Czech Republic\\
$^{37}$ Faculty of Physics, Sofia University, Sofia, Bulgaria\\
$^{38}$ Faculty of Science, P.J.~\v{S}af\'{a}rik University, Ko\v{s}ice, Slovak Republic\\
$^{39}$ Frankfurt Institute for Advanced Studies, Johann Wolfgang Goethe-Universit\"{a}t Frankfurt, Frankfurt, Germany\\
$^{40}$ Fudan University, Shanghai, China\\
$^{41}$ Gangneung-Wonju National University, Gangneung, Republic of Korea\\
$^{42}$ Gauhati University, Department of Physics, Guwahati, India\\
$^{43}$ Helmholtz-Institut f\"{u}r Strahlen- und Kernphysik, Rheinische Friedrich-Wilhelms-Universit\"{a}t Bonn, Bonn, Germany\\
$^{44}$ Helsinki Institute of Physics (HIP), Helsinki, Finland\\
$^{45}$ High Energy Physics Group,  Universidad Aut\'{o}noma de Puebla, Puebla, Mexico\\
$^{46}$ Horia Hulubei National Institute of Physics and Nuclear Engineering, Bucharest, Romania\\
$^{47}$ HUN-REN Wigner Research Centre for Physics, Budapest, Hungary\\
$^{48}$ Indian Institute of Technology Bombay (IIT), Mumbai, India\\
$^{49}$ Indian Institute of Technology Indore, Indore, India\\
$^{50}$ INFN, Laboratori Nazionali di Frascati, Frascati, Italy\\
$^{51}$ INFN, Sezione di Bari, Bari, Italy\\
$^{52}$ INFN, Sezione di Bologna, Bologna, Italy\\
$^{53}$ INFN, Sezione di Cagliari, Cagliari, Italy\\
$^{54}$ INFN, Sezione di Catania, Catania, Italy\\
$^{55}$ INFN, Sezione di Padova, Padova, Italy\\
$^{56}$ INFN, Sezione di Pavia, Pavia, Italy\\
$^{57}$ INFN, Sezione di Torino, Turin, Italy\\
$^{58}$ INFN, Sezione di Trieste, Trieste, Italy\\
$^{59}$ Inha University, Incheon, Republic of Korea\\
$^{60}$ Institute for Gravitational and Subatomic Physics (GRASP), Utrecht University/Nikhef, Utrecht, Netherlands\\
$^{61}$ Institute of Experimental Physics, Slovak Academy of Sciences, Ko\v{s}ice, Slovak Republic\\
$^{62}$ Institute of Physics, Homi Bhabha National Institute, Bhubaneswar, India\\
$^{63}$ Institute of Physics of the Czech Academy of Sciences, Prague, Czech Republic\\
$^{64}$ Institute of Space Science (ISS), Bucharest, Romania\\
$^{65}$ Institut f\"{u}r Kernphysik, Johann Wolfgang Goethe-Universit\"{a}t Frankfurt, Frankfurt, Germany\\
$^{66}$ Instituto de Ciencias Nucleares, Universidad Nacional Aut\'{o}noma de M\'{e}xico, Mexico City, Mexico\\
$^{67}$ Instituto de F\'{i}sica, Universidade Federal do Rio Grande do Sul (UFRGS), Porto Alegre, Brazil\\
$^{68}$ Instituto de F\'{\i}sica, Universidad Nacional Aut\'{o}noma de M\'{e}xico, Mexico City, Mexico\\
$^{69}$ iThemba LABS, National Research Foundation, Somerset West, South Africa\\
$^{70}$ Jeonbuk National University, Jeonju, Republic of Korea\\
$^{71}$ Johann-Wolfgang-Goethe Universit\"{a}t Frankfurt Institut f\"{u}r Informatik, Fachbereich Informatik und Mathematik, Frankfurt, Germany\\
$^{72}$ Korea Institute of Science and Technology Information, Daejeon, Republic of Korea\\
$^{73}$ KTO Karatay University, Konya, Turkey\\
$^{74}$ Laboratoire de Physique Subatomique et de Cosmologie, Universit\'{e} Grenoble-Alpes, CNRS-IN2P3, Grenoble, France\\
$^{75}$ Lawrence Berkeley National Laboratory, Berkeley, California, United States\\
$^{76}$ Lund University Department of Physics, Division of Particle Physics, Lund, Sweden\\
$^{77}$ Nagasaki Institute of Applied Science, Nagasaki, Japan\\
$^{78}$ Nara Women{'}s University (NWU), Nara, Japan\\
$^{79}$ National and Kapodistrian University of Athens, School of Science, Department of Physics , Athens, Greece\\
$^{80}$ National Centre for Nuclear Research, Warsaw, Poland\\
$^{81}$ National Institute of Science Education and Research, Homi Bhabha National Institute, Jatni, India\\
$^{82}$ National Nuclear Research Center, Baku, Azerbaijan\\
$^{83}$ National Research and Innovation Agency - BRIN, Jakarta, Indonesia\\
$^{84}$ Niels Bohr Institute, University of Copenhagen, Copenhagen, Denmark\\
$^{85}$ Nikhef, National institute for subatomic physics, Amsterdam, Netherlands\\
$^{86}$ Nuclear Physics Group, STFC Daresbury Laboratory, Daresbury, United Kingdom\\
$^{87}$ Nuclear Physics Institute of the Czech Academy of Sciences, Husinec-\v{R}e\v{z}, Czech Republic\\
$^{88}$ Oak Ridge National Laboratory, Oak Ridge, Tennessee, United States\\
$^{89}$ Ohio State University, Columbus, Ohio, United States\\
$^{90}$ Physics department, Faculty of science, University of Zagreb, Zagreb, Croatia\\
$^{91}$ Physics Department, Panjab University, Chandigarh, India\\
$^{92}$ Physics Department, University of Jammu, Jammu, India\\
$^{93}$ Physics Program and International Institute for Sustainability with Knotted Chiral Meta Matter (SKCM2), Hiroshima University, Hiroshima, Japan\\
$^{94}$ Physikalisches Institut, Eberhard-Karls-Universit\"{a}t T\"{u}bingen, T\"{u}bingen, Germany\\
$^{95}$ Physikalisches Institut, Ruprecht-Karls-Universit\"{a}t Heidelberg, Heidelberg, Germany\\
$^{96}$ Physik Department, Technische Universit\"{a}t M\"{u}nchen, Munich, Germany\\
$^{97}$ Politecnico di Bari and Sezione INFN, Bari, Italy\\
$^{98}$ Research Division and ExtreMe Matter Institute EMMI, GSI Helmholtzzentrum f\"ur Schwerionenforschung GmbH, Darmstadt, Germany\\
$^{99}$ Saga University, Saga, Japan\\
$^{100}$ Saha Institute of Nuclear Physics, Homi Bhabha National Institute, Kolkata, India\\
$^{101}$ School of Physics and Astronomy, University of Birmingham, Birmingham, United Kingdom\\
$^{102}$ Secci\'{o}n F\'{\i}sica, Departamento de Ciencias, Pontificia Universidad Cat\'{o}lica del Per\'{u}, Lima, Peru\\
$^{103}$ Stefan Meyer Institut f\"{u}r Subatomare Physik (SMI), Vienna, Austria\\
$^{104}$ SUBATECH, IMT Atlantique, Nantes Universit\'{e}, CNRS-IN2P3, Nantes, France\\
$^{105}$ Sungkyunkwan University, Suwon City, Republic of Korea\\
$^{106}$ Suranaree University of Technology, Nakhon Ratchasima, Thailand\\
$^{107}$ Technical University of Ko\v{s}ice, Ko\v{s}ice, Slovak Republic\\
$^{108}$ The Henryk Niewodniczanski Institute of Nuclear Physics, Polish Academy of Sciences, Cracow, Poland\\
$^{109}$ The University of Texas at Austin, Austin, Texas, United States\\
$^{110}$ Universidad Aut\'{o}noma de Sinaloa, Culiac\'{a}n, Mexico\\
$^{111}$ Universidade de S\~{a}o Paulo (USP), S\~{a}o Paulo, Brazil\\
$^{112}$ Universidade Estadual de Campinas (UNICAMP), Campinas, Brazil\\
$^{113}$ Universidade Federal do ABC, Santo Andre, Brazil\\
$^{114}$ Universitatea Nationala de Stiinta si Tehnologie Politehnica Bucuresti, Bucharest, Romania\\
$^{115}$ University of Cape Town, Cape Town, South Africa\\
$^{116}$ University of Derby, Derby, United Kingdom\\
$^{117}$ University of Houston, Houston, Texas, United States\\
$^{118}$ University of Jyv\"{a}skyl\"{a}, Jyv\"{a}skyl\"{a}, Finland\\
$^{119}$ University of Kansas, Lawrence, Kansas, United States\\
$^{120}$ University of Liverpool, Liverpool, United Kingdom\\
$^{121}$ University of Science and Technology of China, Hefei, China\\
$^{122}$ University of South-Eastern Norway, Kongsberg, Norway\\
$^{123}$ University of Tennessee, Knoxville, Tennessee, United States\\
$^{124}$ University of the Witwatersrand, Johannesburg, South Africa\\
$^{125}$ University of Tokyo, Tokyo, Japan\\
$^{126}$ University of Tsukuba, Tsukuba, Japan\\
$^{127}$ Universit\"{a}t M\"{u}nster, Institut f\"{u}r Kernphysik, M\"{u}nster, Germany\\
$^{128}$ Universit\'{e} Clermont Auvergne, CNRS/IN2P3, LPC, Clermont-Ferrand, France\\
$^{129}$ Universit\'{e} de Lyon, CNRS/IN2P3, Institut de Physique des 2 Infinis de Lyon, Lyon, France\\
$^{130}$ Universit\'{e} de Strasbourg, CNRS, IPHC UMR 7178, F-67000 Strasbourg, France, Strasbourg, France\\
$^{131}$ Universit\'{e} Paris-Saclay, Centre d'Etudes de Saclay (CEA), IRFU, D\'{e}partment de Physique Nucl\'{e}aire (DPhN), Saclay, France\\
$^{132}$ Universit\'{e}  Paris-Saclay, CNRS/IN2P3, IJCLab, Orsay, France\\
$^{133}$ Universit\`{a} degli Studi di Foggia, Foggia, Italy\\
$^{134}$ Universit\`{a} del Piemonte Orientale, Vercelli, Italy\\
$^{135}$ Universit\`{a} di Brescia, Brescia, Italy\\
$^{136}$ Variable Energy Cyclotron Centre, Homi Bhabha National Institute, Kolkata, India\\
$^{137}$ Warsaw University of Technology, Warsaw, Poland\\
$^{138}$ Wayne State University, Detroit, Michigan, United States\\
$^{139}$ Yale University, New Haven, Connecticut, United States\\
$^{140}$ Yonsei University, Seoul, Republic of Korea\\
$^{141}$  Zentrum  f\"{u}r Technologie und Transfer (ZTT), Worms, Germany\\
$^{142}$ Affiliated with an institute covered by a cooperation agreement with CERN\\
$^{143}$ Affiliated with an international laboratory covered by a cooperation agreement with CERN.\\

\end{flushleft} 